\documentclass[11pt,a4paper]{article} 
\usepackage{graphicx}
\usepackage{jcappub}

\newcounter{ichi}
\newcounter{ni}
\newcounter{san}
\newcounter{yon}
\makeatletter
\newsavebox{\@parc@ption}
\def\parcaption#1{%
\sbox{\@parc@ption}{\shortstack[l]{#1}}%
>\setbox\@tempboxa\hbox{\csname fnum@\@captype\endcsname}%
\@tempdima\columnwidth \advance\@tempdima-\wd\@tempboxa
\@tempdimb\@tempdima 
\ifdim\wd\@parc@ption>\@tempdimb \@tempdima\@tempdimb
\else\@tempdima\wd\@parc@ption\fi
\sbox{\@tempboxa}{\parbox[t]{\@tempdima}{#1}}%
\caption{\usebox{\@tempboxa}}}
\makeatother

\title{Gamma-ray and neutrino backgrounds as probes of the high-energy universe: hints of cascades, general constraints, and implications for TeV searches}

\author{Kohta Murase$^{1,2}$, John F. Beacom$^{1,2,3}$, and Hajime Takami$^{4}$}

\affiliation{
$^{1}$CCAPP, OSU, 191 W. Woodruff Ave., Columbus, Ohio 43210, USA\\
$^{2}$Department of Physics, OSU, 191 W. Woodruff Ave., Columbus, Ohio 43210, USA\\
$^{3}$Department of Astronomy, OSU, 140 W. 18th Ave., Columbus, Ohio 43210, USA\\
$^{4}$Theory Center, Institute of Particle and Nuclear Studies, KEK, 1-1, Oho, Tsukuba 305-0801, Japan
}

\emailAdd{murase.2@osu.edu, beacom.7@osu.edu, takami@post.kek.jp}

\abstract{
Recent observations of isotropic diffuse backgrounds by \textit{Fermi} and IceCube allow us to get more insight into distant very-high-energy (VHE) and ultra-high-energy (UHE) gamma-ray/neutrino emitters, including cosmic-ray accelerators/sources. 
First, we investigate the contribution of intergalactic cascades induced by gamma-rays and/or cosmic rays (CRs) to the diffuse gamma-ray background (DGB) in view of the latest \textit{Fermi} data.  We identify a possible ``VHE Excess" from the fact that the \textit{Fermi} data are well above expectations for an attenuated power law, and show that cascades induced by VHE gamma rays (above $\sim 10$~TeV) and/or VHECRs (below $\sim 10^{19}$~eV) may significantly contribute to the DGB above $\sim 100$~GeV. 
The relevance of the cascades is also motivated by the intergalactic cascade interpretations of extreme TeV blazars such as 1ES 0229+200, which suggest very hard intrinsic spectra.
This strengthens the importance of future detailed VHE DGB measurements.   
Then, more conservatively, we derive general constraints on the cosmic energy budget of high-energy gamma rays and neutrinos based on recent {\it Fermi} and IceCube observations of extragalactic background radiation.  We demonstrate that these multi-messenger constraints are useful and the neutrino limit is very powerful for VHE/UHE hadronic sources. 
Furthermore, we show the importance of constraints from individual source surveys by future imaging atmospheric Cherenkov telescopes such as Cherenkov Telescope Array, and demonstrate that the cascade hypothesis for the VHE DGB can be tested by searching for distant emitters of cascaded gamma rays. 
}


\begin{document}
\maketitle

\flushbottom

\section{Introduction}
What does the universe look like at energies above 1~TeV per particle?  This is largely unknown, though there are clear hints that this picture will be rich, profound, and may reveal new physics. The key to exploring and deciphering the high-energy universe will be a multi-wavelength and especially multi-messenger approach that synthesizes the information from cosmic rays (CRs), gamma rays, and neutrinos.

CRs are observed at energies up to $\sim {10}^{20}$~eV, but the directions to their sources are scrambled by magnetic fields.  Their flux and spectrum tell us that luminous, high-energy accelerators must exist, which implies the production of high-energy gamma rays and neutrinos. 
High-energy neutrinos from astrophysical sources can uniquely reveal their nature -- for example, the observation of even a few neutrinos indicates that hadrons are being accelerated -- but no high-energy neutrino sources have been detected yet.

On the other hand, luminous gamma-ray sources have been observed up to $\sim 100$~TeV for Milky Way sources and $\sim 10$~TeV for extragalactic sources, but it is not yet known if these are the cosmic-ray production sites.  In the GeV range, the Large Area Telescope (LAT) onboard the {\it Fermi} satellite has detected extragalactic gamma rays from active galactic nuclei (AGN), star-burst galaxies, and gamma-ray bursts (GRBs)~\citep[e.g.,][]{abd+09,abd+10b,abd+11,der12}.  But only a small fraction of the sky has been covered at energies above the typical maximum energy of the \textit{Fermi} gamma-ray telescope, $\sim 300-1000$~GeV, because of the small fields of view of ground-based imaging atmospheric Cherenkov telescopes (IACTs).
  
It will be challenging to observe the most interesting extragalactic sources because very-high-energy (VHE; $>0.1$~TeV) gamma rays from distant sources interact with the extragalactic background light (EBL) and the cosmic microwave background (CMB) via the electron-positron pair production process, so the original signals are depleted.  
However, electron-positron pairs generated by multi-TeV and higher-energy photons up-scatter CMB and EBL photons via the inverse-Compton (IC) scattering process to produce secondary, cascade gamma rays, which fall typically in the energy range of \textit{Fermi}.  

Such VHE gamma rays can be produced via both leptonic and hadronic mechanisms inside astrophysical sources.  For example, some blazars, which are thought to be AGN seen by an on-axis observer, show $\gtrsim$~TeV emissions that can be explained by either leptonic or hadronic mechanisms~\citep[e.g.,][]{abd+11b}. 
If CR ions are accelerated to sufficiently high energies, VHE or even ultra-high-energy (UHE; $>{10}^{19}$~eV) gamma rays are emitted from the sources via photonuclear and/or hadronuclear reactions and subsequent cascades inside the sources. 
Sufficiently high-energy CRs also induce intergalactic cascades after they escape from the sources, as often discussed in the context of cosmogenic neutrinos~\citep[e.g.,][]{bz69,tak+09,kot+10}.
 
These intergalactic electromagnetic cascades are important for using gamma rays as a probe of the VHE/UHE universe.  Together, high-energy gamma-ray and neutrino probes can reveal the existence and nature of high-energy sources, including astrophysical (e.g., CR accelerators) and exotic (e.g., dark matter annihilation or decay) sources.

The diffuse gamma-ray background (DGB) was measured by LAT above 200~MeV~\cite{abd+10a}, and it turned out that its flux at $\sim 1-10$~GeV is much lower than that obtained by EGRET~\cite{sre+98,str+04}. 
The origin of the DGB is still under debate, and many theoretical attempts have been made in order to explain it (see~\cite{der12} and references therein).  However, none are universally accepted although guaranteed contributors such as undetected AGN and star-burst galaxy populations are widely accepted.  All can be argued to be only a subdominant component of the observed background, and it may be that even their sum is insufficient to explain the measurements.  

This presents a special opportunity to explore the contribution from higher energies, cascaded down into the \textit{Fermi} range.  Generally speaking, intergalactic cascade contributions can be crucial for injections at sufficiently high energies~\cite{ca97,cai07,ven10}.  This is even the case when secondary pair production is more effective due to cascades induced by UHE gamma rays or VHECRs/UHECRs.  In preliminary \textit{Fermi} data at energies above $\sim 100$~GeV, as we show, there may be hints of such a cascade contribution.  

In any case, we can develop strong limits on such high-energy contributions by requiring that the calculated spectra do not exceed the \textit{Fermi} data at lower energies around 100~GeV, where the cascade contribution would peak. 
It is difficult for IACTs to measure the VHE DGB.  However, thanks to their good sensitivities, individual gamma-ray sources can be seen via coordinated observations or systematic surveys with better photon statistics.  This enables us to connect the origin of VHE DGB to TeV observations.  

On the neutrino side, there is good evidence from the CR and gamma-ray data that present neutrino detectors are nearly at the required sensitivity for first detections.  Interestingly, for certain hadronic models, the neutrino limits on high-energy injections are considerably stronger than the associated gamma-ray limits.   
At energies in the GeV-TeV range, detectors like IceCube are limited in the sensitivity because of the atmospheric neutrino background; this means that the flux limits improve slowly with increased exposure.  In contrast, in the $\gtrsim$~PeV range, IceCube has so far not detected limiting backgrounds, so that the flux limits are already strong and should improve linearly with time.  
This gives another special opportunity.

In this work, we study the contribution of electromagnetic cascades induced by VHE/UHE gamma rays and VHECRs, with detailed numerical calculations performed in ~\cite{mur+12}.  
We show that a component with cascades may start to dominate over the usual attenuated gamma-ray component above $\sim 100$~GeV, and this may be compatible with preliminary \textit{Fermi} data (Section 2).  
Then, in Section 3, we use the DGB to place constraints on the energy budget of gamma rays, and the results are compared to IceCube constraints on the neutrino energy budget (Section 3).  Our limits are very general, and they complement specific limits for VHECR/UHECR sources. 
In Section 4, we discuss implications for future surveys by next-generation IACTs~\cite{hh10} such as the Cherenkov Telescope Array (CTA)~\cite{act+11}, and estimate expected constraints on populations of VHE/UHE gamma-ray and VHECR sources.  
Finally, based on the results obtained in Sections 3 and 4, we demonstrate that future searches for individual distant sources provide an important diagnostic to test the cascade hypothesis for the VHE DGB. 
In this paper, we adopt the $\Lambda$CDM model with $H_0 \equiv 100h =71~{\rm km}~{\rm s}^{-1}~{\rm Mpc}^{-1}$, $\Omega_m=0.3$ and $\Omega_{\Lambda}=0.7$.

\section{Cascade contributions to the diffuse gamma-ray background}
LAT extracts the DGB up to $\sim 100$ GeV by subtracting contributions of resolved individual sources (mainly AGN for extragalactic sources) and Galactic components seen at high latitudes~\cite{abd+10a}.  
The DGB observed by LAT is significantly lower than that obtained by EGRET above the GeV range~\cite{sre+98,str+04}.  Its spectrum can be fitted by a power law with a differential spectral index $\alpha=2.41 \pm 0.05$ and intensity integrated over energies above 100~MeV, $I_{\gamma} (>100~{\rm MeV}) = (1.03 \pm 0.17) \times {10}^{-5}~{\rm cm}^{-2}~{\rm s}^{-1}~{\rm sr}^{-1}$, where the error is systematics-dominated~\cite{abd+10a}.  Throughout the paper, we use  
\begin{equation}
E_{\gamma}^2 \Phi_{\gamma} = 8.55 \times {10}^{-8}~{\rm GeV}~{\rm cm}^{-2}~{\rm s}^{-1}~{\rm sr}^{-1} ~{(E_{\gamma}/100~{\rm GeV})}^{-0.41}, 
\end{equation}
for the measured differential background intensity (in the unit of particles per energy, area, time and solid angle) $\Phi_\gamma$, where the normalization is given by $I_\gamma = \int d E_\gamma \, \Phi_\gamma$.  Recently, the DGB up to $\sim 600$~GeV has also been shown as a preliminary result~\footnote{See the presentation by M. Ackermann on behalf of the {\it Fermi} collaboration: http://agenda.albanova.se/conferenceDisplay.py?confId=2600.}, and its spectrum looks basically consistent with the Eq.~(2.1).  

Figures~1 and 2 show the \textit{Fermi} data, which agree well with the power-law spectrum expressed by Eq.~(2.1).  However, if the sources are extragalactic, then EBL attenuation is inevitable.  The simple power-law spectrum of primary gamma rays emitted from the sources, with $\alpha = 2.41$, provides a quite reasonable fit at low energies, but falls well below the data above $\sim 100$~GeV in both the figures after the EBL attenuation is taken into account.  
This means that it is crucially important to settle \textit{Fermi} data analyses of the VHE DGB and extend VHE DGB measurements to further higher energies.  In showing an example of the EBL-attenuated power-law spectrum, the low-IR model of Ref.~\cite{kne+04} is adopted, which is one of the conservative models.  The attenuation is more severe for many other models, e.g., the best-fit model (as demonstrated in Figures~1 and 2) and/or stellar-UV model in the same reference; see Ref.~\cite{fin+10} and references therein for other models.  This conclusion would not change by due to uncertainties on the EBL when most of the gamma rays are injected at $z \sim 1-2$, as expected for evolution of the gamma-ray emissivity of many astrophysical objects including blazars~\citep[e.g.,][]{it09,dr10} and star-burst/star-forming galaxies~\citep[e.g.,][]{fie+10,mak+11}.  The quantitative results also depend redshift evolution models; the excess can be weaker for no or negative redshift evolution models with lower EBL models, though it is still identified for the EBL models used here, as shown in Figure~2.  Note that the redshift evolution model by Ref.~\cite{hb06} for star formation evolution is adopted throughout the paper.  Thus, one may think that the ``VHE Excess" is  an interesting issue for the extragalactic origin of the DGB. 

In this work, we assume that sources responsible for the DGB are cosmic.
In deriving the DGB, the Galactic contribution is removed based on an emission model as well as the emission from the Sun and the CR residual background.  Hence, \textit{unaccounted-for} Galactic components (e.g., millisecond pulsars~\cite{fl10,sie+11}, dark matter~\cite{mb12}, debris at the outer frontier of the solar system~\cite{mp09}, IC scattering of solar photons with local CRs~\cite{mos+06}) and/or \textit{unaccounted-for} CR-induced detector backgrounds could potentially contribute.  In particular, an unresolved population of millisecond pulsars is difficult to explain the VHE DGB since its typical spectrum shows a cutoff around a few~$\times$~GeV~\cite{der12}. 

\begin{figure*}[bt]
\begin{minipage}{0.49\linewidth}
\begin{center}
\includegraphics[width=\linewidth]{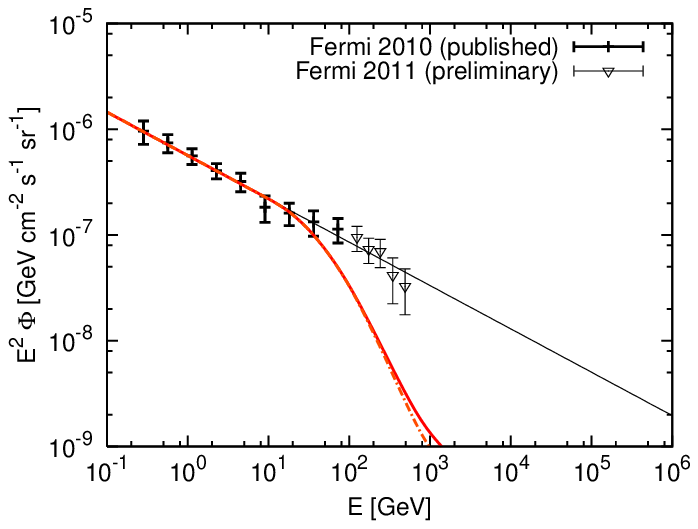}
\caption{Published DGB data~\cite{abd+10a} and preliminary DGB data. 
Examples of DGBs fitted by power laws without (thin curve)~\cite{abd+10a} or with (thick curves) EBL attenuation are also shown, where we use $a=1$ and $\alpha=2.41$ defined in Eq.~(2.7).  
We identify the ``VHE Excess" relative to the EBL-attenuated simple power law.  Star formation evolution~\cite{hb06} is assumed.  The solid and the dot-dashed curves are for the low-IR model and the best-fit model of Ref.~\cite{kne+04}, respectively. 
} 
\end{center}
\end{minipage}
\begin{minipage}{.1\linewidth}
\end{minipage}
\begin{minipage}{0.49\linewidth}
\begin{center}
\includegraphics[width=\linewidth]{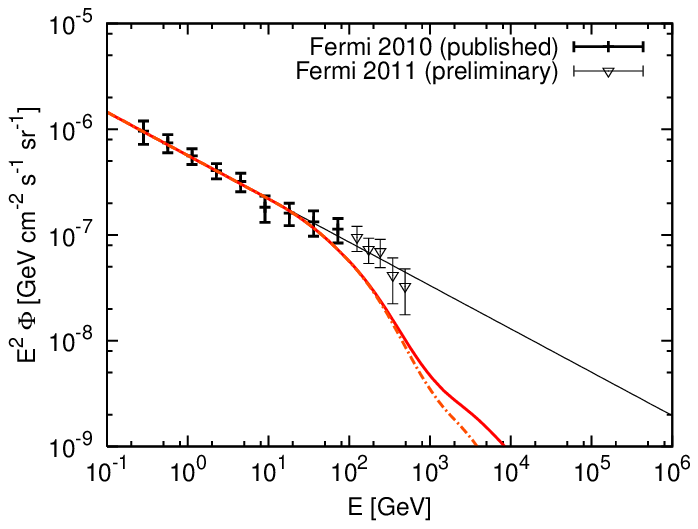}
\caption{The same as Figure~1, but no redshift evolution is assumed. 
\newline \,  \newline \, \newline \, \newline \, \newline \, \newline \, \newline \, \newline 
}
\end{center}
\end{minipage}
\end{figure*}

Because the EBL attenuation becomes relevant in the energy range of the preliminary data, the simplest power-law extension of gamma-ray emission spectra from low energies may start to fail above $\sim 100$~GeV~\citep[c.f.][]{ino11}.  If this excess is correct, though future careful investigations are necessary, the VHE DGB requires introducing an additional hard component.   
Then, one of the natural questions is how hard this additional component is.  If the spectral indices are steep enough, cascades would not play an important role.  Note that this hard component itself does not explain the DGB at $\sim 1$~GeV.  Figure~3 suggests that, when we consider only EBL attenuation, very hard effective photon indices of the primary gamma-ray spectrum ($\alpha \lesssim 2$) are required to fit the preliminary \textit{Fermi} data.  Although the ``VHE Excess" seems explained at first glance, neglecting cascades is physically incorrect for such hard spectra.  As we show below, for adequately luminous and sufficiently hard primary spectra of extragalactic gamma rays, \textit{one must take into account intergalactic cascade contributions in the relevant range when the primary gamma-ray spectrum extends to $\gtrsim 10$~TeV.}   
Hereafter, we consider potential roles and impacts of cascades in detail.         

VHE gamma rays are produced in sources via leptonic processes (e.g., synchrotron self-Compton and external IC processes) and/or hadronic processes (e.g., $p \gamma$ processes).  UHE gamma rays can be generated only via hadronic processes.  In addition, they can be produced by sufficiently high-energy CRs escaping from the source, through hadronuclear (e.g., $pp$) and photonuclear (e.g., $p \gamma$) collisions.  VHE/UHE gamma rays from distant sources cannot avoid pair creation with the EBL and CMB, depending on their energies, and will be cascaded down to GeV-TeV energies.  Whatever the origin of the DGB is, if the cascade is sufficiently developed, it has a near-universal form~\citep[e.g.,][]{bs75,ca97} as
\begin{equation}
G_{E_{\gamma}} \propto 
\left\{ \begin{array}{ll}
{(E_{\gamma}/E_{\gamma}^{\rm br})}^{-1/2}
& \mbox{($E_\gamma \leq E_{\gamma}^{\rm br}$)}\\
{(E_{\gamma}/E_{\gamma}^{\rm br})}^{1-\beta} 
& \mbox{($E_{\gamma}^{\rm br} < E_{\gamma} \leq E_{\gamma}^{\rm cut}$)},
\end{array} \right. 
\end{equation} 
where $E_\gamma G_{E_{\gamma}}$ represents the shape of the energy spectrum of cascaded gamma rays and its normalization is determined by $\int d E_\gamma \, G_{E_{\gamma}} = 1$.  Here, $E_{\gamma}^{\rm cut}$ is the energy where the suppression due to pair creation occurs, $\beta$ is typically $\sim 2$, and $E_{\gamma}^{\rm br} \approx (4/3) {({E'}_\gamma^{\rm cut}/2 m_e c^2)}^2 \varepsilon_{\rm CMB} \simeq 0.034~{\rm GeV}~{(E_{\gamma}^{\rm cut}/0.1~{\rm TeV})}^2 {((1+z)/2)}^{2}$ is the break energy corresponding to $E_{\gamma}^{\rm cut}$, where $\varepsilon_{\rm CMB}$ is the typical CMB energy.  
We are interested in cases of ${E}_{\gamma}^{\rm max} > {E}_\gamma^{\rm cut}$.  Then, the spectral shape above $\sim E_{\gamma}^{\rm cut}$ and below $\sim {\rm min}[{E}_\gamma^{\rm max}/2,(4/3) {({E'}_\gamma^{\rm max}/2 m_e c^2)}^2 \varepsilon_{\rm CMB}]$ depends on intergalactic gamma-ray injection processes~\footnote{For examples of the cascade DGB spectrum for ${E'}_{\gamma}^{\rm max} \sim 1-10$~TeV, see Murase et al.~\cite{mur+07}.}.  For the gamma-ray-induced cascade, when the cascade mainly occurs in the Thomson regime, we may roughly expect $e^{-\tau_{\gamma \gamma}}$ for distant sources with $d > \lambda_{\gamma \gamma}$, where $\tau_{\gamma \gamma} (E_\gamma,z) \sim d(z)/\lambda_{\gamma \gamma} (E_\gamma)$ is the optical depth for the pair creation.  
On the other hand, for the gamma-ray-induced cascade in the Klein-Nishina regime and VHECR-induced cascade, pairs are more continuously supplied so that the high-energy spectral shape is more like $(1-e^{-\tau_{\gamma \gamma}})/\tau_{\gamma \gamma}$ as long as the gamma-ray and pair injection length (e.g., the Bether-Heitler energy loss length $\lambda_{\rm BH}$ for the CR-induced cascade) is longer than $d$~\citep[e.g.,][]{ess+11}.  The latter is analogous to the solution of the radiative transfer equation for the uniform slab model, where the absorbing and emitting regions are co-spatial.

In order to obtain detailed spectra of cascaded gamma rays, for this work, we perform numerical calculations of the gamma-ray cascade by solving Boltzmann equations taking into account pair creation, IC emission, synchrotron emission, and adiabatic loss~\citep[for details, see][]{mur12,mb12,mur+12}.  The background flux can be evaluated from the following formula~\citep[e.g.,][]{ste+93,mur+07},
\begin{equation}
E^2 \Phi = \frac{c}{4 \pi} \int dz  \,\,\, \left| \frac{d t}{d z} \right| E^2 \frac{d \dot{n}}{d E} (z),
\end{equation}
where $E^2 (d \dot{n}/d E)$ is the observed energy spectrum of the energy output rate resulting from injections at $z$ (in unit of energy per time per comoving volume) and $dt/dz$ is given by
\begin{equation}
\left| \frac{d t}{d z} \right| = \frac{1}{H_0 (1+z) \sqrt{(1+z)^3 \Omega_m+\Omega_\Lambda}}
\end{equation}
in the flat $\Lambda$CDM universe. 
It is convenient to introduce the energy budget $Q (z)$. In the case of primary gamma-ray injections, the total gamma-ray energy budget at $z$ is  
\begin{equation}
Q_{\gamma} (z) = \int d E_{\gamma}^\prime \,\,\, E_\gamma^\prime \frac{d \dot{n}_{\gamma}^\prime}{d E_{\gamma}^\prime} \equiv \int d E_\gamma^\prime \,\,\, Q_{E_\gamma^\prime} 
\end{equation} 
and the local energy budget is defined as $Q_\gamma \equiv Q_\gamma (z=0)$. 

Examples of resulting spectra of cascaded gamma rays for the DGB are shown in Figure~4.  As expected in Eq.~(2.2), their detailed shape is not sensitive to the injection spectrum as long as the gamma-ray injection energy is high enough. 
In the case of star formation evolution, the GeV gamma-ray emission mainly comes from sources around $z \sim 1-2$, so the flux is larger by a factor of $\sim 5$.  The enhancement is even more significant if redshift evolution is much faster~\citep[e.g.,][]{yk07,gel+11}.  On the other hand, the TeV emission mainly comes from nearby sources due to EBL absorption, so it does not have a strong dependence on redshift evolution models for the same gamma-ray energy budget.

\begin{figure*}[bt]
\begin{minipage}{0.49\linewidth}
\begin{center}
\includegraphics[width=\linewidth]{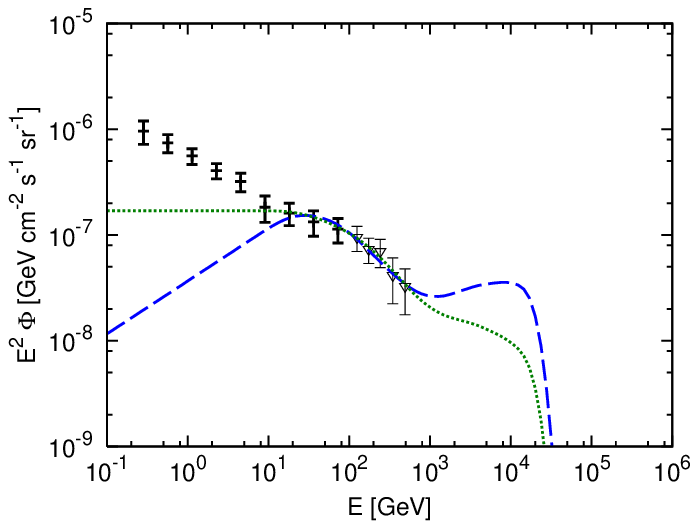}
\caption{
Examples of VHE DGBs fitted by EBL-attenuated power laws.  We use $a=0.065$ and $\alpha=1.5$ for star formation evolution (dashed curve) or $a=0.30$ and $\alpha=2.0$ for no redshift evolution (dotted curve), with the low-IR model of Ref.~\cite{kne+04}.  
These cases are unrealistic unless ${E'}_{\gamma}^{\rm max}$ is low enough, because cascades are neglected (c.f. Figures~5-7).  
} 
\end{center}
\end{minipage}
\begin{minipage}{.1\linewidth}
\end{minipage}
\begin{minipage}{0.49\linewidth}
\begin{center}
\includegraphics[width=\linewidth]{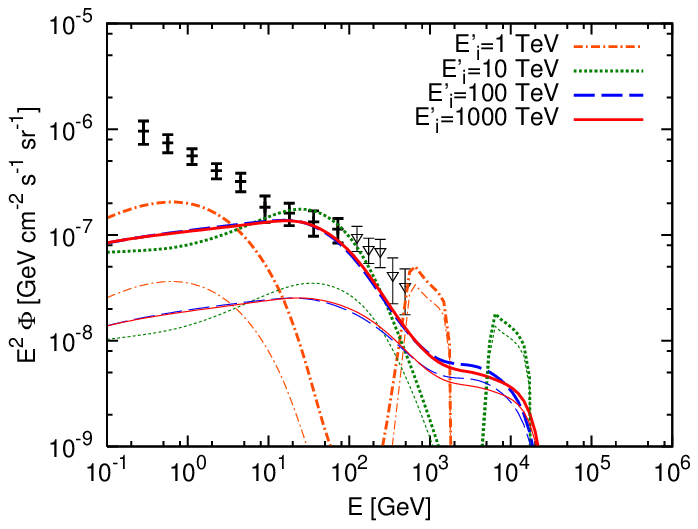}
\caption{Examples of DGBs from cascaded gamma rays for a near-mono-energetic gamma-ray spectrum injected over a half decade in logarithmic energy (with a central value, $E'_i$).  We use $Q_\gamma = {10}^{45}~{\rm erg}~{\rm Mpc}^{-3}~{\rm yr}^{-1}$.   
The thick and thin curves are for star formation evolution and no redshift evolution, respectively.  
\newline
}
\end{center}
\end{minipage}
\end{figure*}

The origin of the DGB has been a mystery.  Radio-loud AGN including blazars are the most popular sources~\citep[e.g.,][]{ste+93,it09,aba+11,aba+12,yan+12}, whereas star-forming and star-burst galaxies may also give a significant contribution to the DGB~\cite{pf02,lw06,tho+06,fie+10,mak+11}.  Many models predict that the typical photon index of the DGB is $\alpha > 2$~\cite{tho+06,it09,mak+11,aba+11,yan+12}.  Although the EBL model has some uncertainties~\cite[see][and references therein]{fin+10}, including attenuation due to the EBL should lead to suppression above $\sim 100$~GeV (Figures~1 and 2).  Obviously, it is crucial whether the measured VHE DGB is consistent with a simple extrapolation of the DGB from the GeV range or not.  
If the ``VHE Excess" is confirmed, however, it may suggest the existence of a distinct component above $\sim 100$~GeV.  Sources responsible for this component can be different from sources contributing the $\sim 1$~GeV background (e.g., BL Lac objects for the VHE DGB and some combination of other sources for $\sim 1$~GeV).  Or, the effective primary gamma-ray spectrum of dominant sources (e.g., blazars) may be concave if there are many very hard TeV sources that cannot be seen by \textit{Fermi}.  In any case, it will increase the importance of the cascade contributions.   

In this paper, though it is also possible to invoke the dark matter origin~\cite{mb12}, we consider astrophysical interpretations of the VHE DGB.  Generally speaking, the origin of primary VHE/UHE gamma rays can be either leptonic or hadronic, or injections occur during propagation of VHECRs.  
In extragalactic scenarios, one straightforward way to compensate for the severe EBL attenuation is to think that the primary gamma-ray spectrum hardens in the TeV range.  For example, the effective spectrum of primary gamma rays escaping from VHE sources may be concave, and we here point out that the VHE DGB could be reconciled if the primary gamma-ray spectrum is not a simple power law but has an appropriate bump in the $\sim 0.1-1$~TeV range.  Or, we can exploit the EBL-attenuated power-law spectrum for primary gamma-ray injections with ${E'}_\gamma^{\rm max}\lesssim 10$~TeV.  However, when the primary spectrum extends to higher energies with an effective photon index comparable to or harder than $2$ (as expressed by the dashed and dotted curves in Figure~3), one cannot neglect the cascade contribution.  In this case, cascades also give us a reasonable upper limit on the VHE DGB for some kinds of primary gamma-ray spectra.  Hereafter, we focus on the latter case where the intergalactic cascade is important.

\subsection{Cascades induced by primary VHE/UHE gamma rays}
Cascade signatures might indeed exist in VHE spectra of observed individual sources.  For example, it seems that a fraction of blazars indeed have a very hard spectrum, as observed for 1ES 0229+200~\cite{aha+07b} and 1ES 1101-232~\cite{aha+07}, and it might be the case especially at low luminosities.  Interestingly, such a class of extreme blazars can be explained by the intergalactic cascade emissions~\citep[see, e.g.,][and references therein]{ek10,ess+11,mur+12,raz+12}.  
Motivated by these recent studies, we consider a hard primary gamma-ray spectrum with a power-law form as
\begin{equation}
E_\gamma^\prime Q_{E_\gamma^\prime} \propto {(E_\gamma^\prime)}^{2-s} \,\,\, (E_\gamma^\prime \leq {{E'}_{\gamma}^{\rm max}}),
\end{equation}
where the normalization is determined by Eq.~(2.5).  One sees that basic features of the intergalactic cascade spectrum are the same as those shown in Figure~4, and results of the spectra of cascaded gamma rays for $s=1.5$ and $s=2$ are shown in Figures~5, 6 and 7.  
As indicated from Figure~4, the spectrum of cascaded gamma rays becomes almost universal for sufficiently high-energy gamma-ray injections.  Note that, for the near-mono-energetic injection in the $\sim 0.1-10$~TeV range, given that its cascade component does not exceed the published \textit{Fermi} data, its attenuated component can be higher than the near-universal cascade spectrum that is seen for ${E'}_{\gamma}^i=100$~TeV and ${E'}_{\gamma}^i=1000$~TeV.      

In order to model the EBL-attenuated power-law component, we adopt the following parameterization, 
\begin{equation}
E_{\gamma}^2 \Phi_{\gamma} = 8.55 a \times {10}^{-8}~{\rm GeV}~{\rm cm}^{-2}~{\rm s}^{-1}~{\rm sr}^{-1}~{(E_{\gamma}/100~{\rm GeV})}^{2-\alpha} e^{-\tau_{\gamma \gamma}^{\rm eff}(E_\gamma,\alpha)},
\end{equation}
where $\tau_{\gamma}^{\rm eff}(E_\gamma,\alpha)$ is the effective intergalactic optical depth that is easily calculated by Eq.~(2.3), using $\tau_{\gamma \gamma} (E_\gamma, z)$, as  
\begin{equation}
e^{-\tau_{\gamma \gamma}^{\rm eff}(E_\gamma,\alpha)} \propto \int dz \,\,\, \left| \frac{d t}{d z} \right| \frac{Q_\gamma (z)}{{(1+z)}^{\alpha-1}} e^{-\tau_{\gamma \gamma} (E_\gamma,z)}. 
\end{equation}
Here the normalization is set by requiring that Eq.~(2.7) corresponds to Eq.~(2.1) in the transparent limit.   
Note that cascade contributions from this power-law component are insignificant when $\alpha$ is sufficiently larger than 2~\citep[e.g.,][]{it09,ino11}. 
One should keep in mind that the real situation can be complicated by various astrophysical contributions especially from blazars and star-forming/star-burst galaxies~\citep[see a review][and references therein]{der12}.  But taking them into account depends on many source-dependent details including the luminosity function, spectral energy distribution, redshift evolution and so on.  To avoid such complexities and uncertainties, we focus on this simple case.  Although such detailed modeling, including fitting procedures, should be done when the VHE DGB is precisely measured, our approach here is enough for the current purpose, which is to show the likely importance of intergalactic cascade contributions and to demonstrate the potential impacts of measuring the VHE DGB.    
 
\begin{figure*}[tb]
\begin{minipage}{0.49\linewidth}
\begin{center}
\includegraphics[width=\linewidth]{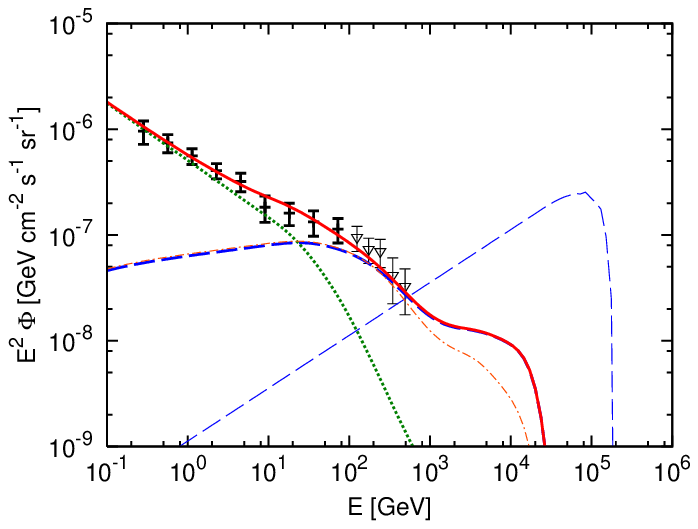}
\caption{
An example of how a cascade component could help explain the measured VHE DGB.  A low-energy component, with EBL attenuation, is shown by the dotted curve, where $a=0.9$ and $\alpha=2.54$ are assumed. 
A primary gamma-ray component is shown by the thin dashed curve, where the effective photon index is set to $s=1.5$ with ${E'}_{\rm max}={10}^{2.25}$~TeV and $Q_{\gamma}=2 \times {10}^{45}~{\rm erg}~{\rm Mpc}^{-3}~{\rm yr}^{-1}$, and no redshift evolution are assumed.  
The resulting component with cascades is shown by the thick dashed curve, and the solid curve is the sum of the two components.  The low-IR model of Ref.~\cite{kne+04} is used, but the cascaded gamma-ray spectrum for their best-fit model (dot-dashed curve) is also shown. 
}
\end{center}
\end{minipage}
\begin{minipage}{.05\linewidth}
\end{minipage}
\begin{minipage}{0.49\linewidth}
\begin{center}
\includegraphics[width=\linewidth]{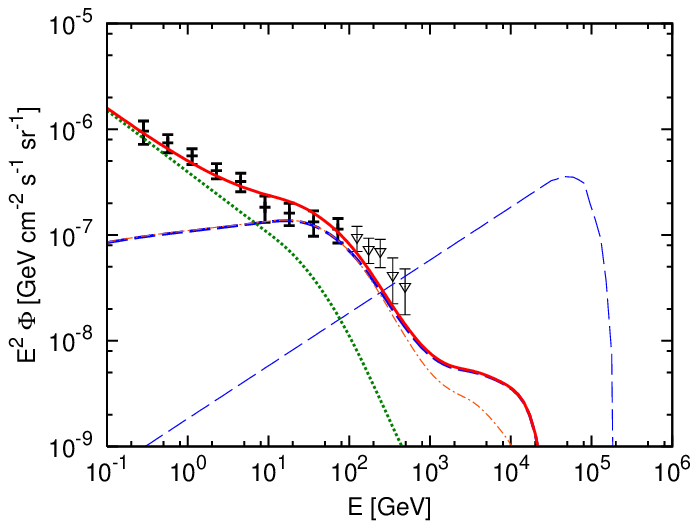}
\caption{The same as Figure~5, but $Q_{\gamma}=6 \times {10}^{44}~{\rm erg}~{\rm Mpc}^{-3}~{\rm yr}^{-1}$ with star formation evolution~\cite{hb06}. 
For the EBL-attenuated power-law component, $a=0.7$ and $\alpha=2.58$ are assumed.  
\newline \, \newline \, \newline \, \newline \, \newline \, \newline \, \newline \, \newline \, \newline \, \newline \, 
}
\end{center}
\end{minipage}
\end{figure*}
 
In Figure~5, it is demonstrated that the contribution with cascades could indeed explain the VHE DGB and that the component with cascades can dominate over the EBL-attenuated power-law component.  
The spectrum of secondary gamma rays typically leads to a bump feature around 100~GeV, which looks consistent with an apparent tendency of the DGB.  Also, this suggests that the published data below 100~GeV correspond to the most important energies for setting general cascade bounds that are discussed in Section~3.  At higher energies, the cascade spectrum for hard photon indices also gives us a conservative maximum contribution to the VHE DGB for power-law injections of primary gamma rays with ${E'}_{\gamma}^{\rm max} \gtrsim 10$~TeV or $s>2$, although this does not generally hold for, e.g., a spiky primary spectrum in the $\sim 0.1-10$~TeV range (cf. Figure~4), which can be expected in the very heavy dark matter scenario~\cite{mb12}.   
One also sees that the slope above $\sim 100$~GeV is steeper for faster redshift evolution models (Figure~6).  Hence, if the preliminary data points are indeed correct, the fit with cascades would be better in the no redshift evolution case.  In other words, the cascade spectrum in the no redshift evolution case gives a reasonable upper limit on the DGB above $\sim 100$~GeV.  
The tell-tale shape of the VHE DGB is best revealed in the range of the preliminary data above $\sim 100$~GeV, so precise measurements of the VHE DGB can exclude sufficiently fast redshift evolution models given that the EBL is well determined.  Also, if the finalized VHE DGB data are higher than the theoretical predictions, Galactic contributions may be relevant~\cite{mb12}.  
For comparison, we also show an example of higher EBL models, using the best-fit model of Ref.~\cite{kne+04}.  Choosing different EBL models affects the quantitative results, but one sees that basic results are not changed in this case.  Throughout this paper, we use the low-IR model of Ref.~\cite{kne+04}, which is one of the conservative EBL models.  In principle, Figures~5 and 6 suggest that measuring the VHE DGB (at $\gtrsim 0.5$~TeV) allows us to test the cascade hypothesis and other extragalactic scenarios if the EBL is determined.  Or, given the cascade hypothesis, the EBL can be constrained.    

The spectral index of $s=1.5$ may be extreme, and sources with such very hard spectra may be rather rare.  More conservatively, in Figure~7, we consider the case of $s=2.0$, where $s$ should be regarded as an effective photon index, i.e., the index averaged over sources with various photon indices.  For example, some combination of several sources such as radio galaxies and star-forming/star-burst galaxies could be responsible for the DGB at $\sim 1$~GeV, whereas unresolved BL Lac objects (including ones with very hard primary spectra) might give a dominant contribution to the VHE DGB.  Figure~7 suggests that the contribution with cascades is not negligible for hard spectral indices used in Figure~3.  Taking only the attenuation into account is not physically valid unless the maximum energy is low enough.      

In Figure~8, we show the results for the UHE gamma-ray injection. 
Such UHE gamma rays should have a hadronic origin and can be produced via photomeson production inside the source~\citep[e.g.,][]{ad03,mur09,der+12}. For example, in jets of GRBs and AGN, high-energy protons and neutrons interact with target photons, leading to UHE neutrinos and gamma rays.  If the synchrotron self-absorption or Rayleigh-Jeans portion of the black-body spectrum is relevant, UHE gamma rays that mainly come from neutral pion decay may escape into intergalactic space~\footnote{Magnetized environments including structured regions (galaxy clusters and filaments) and the immediate environment (e.g., galaxies and dust tori) also lead to \textit{synchrotron pair halo/echo emissions}, as suggested by Murase~\cite{mur12}.}.  
In Figure 8, we consider a near-mono-energetic UHE gamma-ray injection around ${E'}_{\gamma}^{\rm max}=10$~EeV, and the height of the energy flux of primary gamma rays is comparable to that of cascaded gamma rays (because the total electromagnetic energy amount should be conserved before and after cascades).  As expected, the cascade spectrum is not sensitive to primary spectra of UHE gamma rays, so the results in the GeV-TeV range are similar to Figures~5 and 7.  However, the effective energy loss length of UHE gamma rays is $\sim 10-100$~Mpc because of the leading-particle effect~\cite{mur09,mur12}, so the cascade spectrum shown in Figure~8 is slightly harder due to longer-lasting injections during the Klein-Nishina cascade.  Note that these cases are shown to demonstrate the near-universality of the cascade spectrum.  UHE gamma-ray sources accompanying VHE/UHE neutrinos are not expected to contribute to the DGB though individual sources can be seen (see next sections).  

\begin{figure*}[bt]
\begin{minipage}{0.49\linewidth}
\begin{center}
\includegraphics[width=\linewidth]{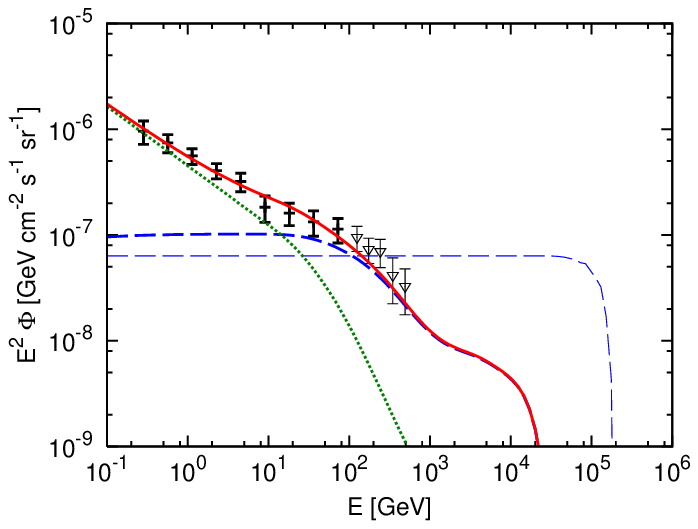}
\caption{
The same as Figure~5, but $s=2$ is used.  For the EBL-attenuated power-law component, $a=0.8$ and $\alpha=2.56$ are assumed.  
\newline 
}
\end{center}
\end{minipage}
\begin{minipage}{.05\linewidth}
\end{minipage}
\begin{minipage}{0.49\linewidth}
\begin{center}
\includegraphics[width=\linewidth]{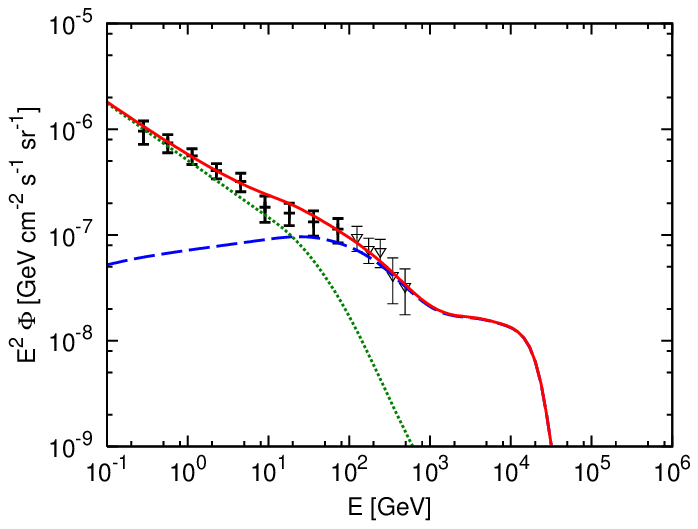}
\caption{
The same as Figure~5, but the near-mono-energetic UHE gamma-ray injection over a half decade in logarithmic energy (with the central value of $E'_{\rm max}=10$~EeV) is assumed. 
}
\end{center}
\end{minipage}
\end{figure*}

\subsection{Cascades induced by primary VHECRs}
Not only VHE/UHE gamma rays but also VHECRs/UHECRs can induce intergalactic cascades. Sufficiently high-energy CRs interact with CMB and EBL photons via the photomeson and Bethe-Heitler production processes, and injected VHE/UHE gamma rays and pairs play an important role in the resulting cascade.  
The processes have been discussed in the literature of UHECRs whose energies reach $\sim {10}^{20}$~eV~\citep[e.g.,][]{ber+11,da11,gel+11}.  In this work, we rather focus on VHECR accelerators, where protons can be accelerated only up to $\sim {10}^{19}$~eV.  
Various sources can be accelerators of VHECRs in this energy range. For example, in the standard blazar model, BL Lac objects and Fanaroff-Riley (FR) I galaxies can typically accelerate protons up to $\sim {10}^{19}$~eV~\cite{mur+12}.  Hypernovae are also potential VHECR accelerators, and low-luminosity GRBs accompanying relativistic ejecta may even produce UHECRs~\cite{mur+08b}.  Accretion and merger shocks in clusters of galaxies have also been suggested as the origin of CRs below the ankle~\cite{mur+08a}. 

In the energy range below the ankle, the allowed CR energy budget is larger than the UHECR energy budget~\citep[e.g.,][]{ber+06,mur+08a,da11}.  Also, importantly, the Bethe-Heitler process is more relevant than photomeson production.  Since this process itself does not include neutrino production, neutrino constraints, which are discussed in the next section, can also be avoided.  
We consider the following proton spectrum,
\begin{equation}
E_p^\prime Q_{E_p^\prime} \propto {(E_p^\prime)}^{2-q} \,\,\, (E_p^\prime \leq {{E'}_p^{\rm max}}),
\end{equation}
where the proton normalization is set by $Q_{\rm vhecr}=\int_{{10}^{18}~{\rm eV}}^{{E'}_p^{\rm max}} d E' \, Q_{E'}$, where the proton maximum energy is set to ${E'}_p^{\rm max}={10}^{19}$~eV throughout this paper.  Note that the total CR energy budget $Q_{\rm cr}$ is larger than the VHECR energy budget $Q_{\rm vhecr}$.  We use $Q_{\rm vhecr}$ that is more directly related to CR observations.  Low-energy CRs do not interact with photons either by the photomeson production or the Bethe-Heitler process, and $Q_{\rm cr}/Q_{\rm vhecr}$ depends on the extrapolation to lower energies.  
For demonstration, we adopt $q=2.6$, motivated by extragalactic scenarios explaining the observed VHECRs below the ankle~\citep[e.g.,][]{ber+06,mur+08a}.  
Unless one considers redshift evolution models faster than star formation evolution, such steep indices are required for the VHECR-induced cascade component to contribute to the DGB~\citep[e.g.,][]{da11,gel+11}, where the main process should be the Bethe-Heitler process rather than the photomeson production.  In other words, as seen later, the UHECR energy budget is typically smaller than the required gamma-ray energy budget.    

Our results are shown in Figures~9 and 10, where it is shown that the VHECR-induced cascade contribution could marginally explain the VHE DGB and the cascade component can dominate over the EBL-attenuated power-law component.  
Because the Bethe-Heitler process can inject pairs over $\lambda_{\rm BH} \gg \lambda_{\gamma \gamma}$, the VHECR-induced cascade spectrum is slightly harder than the gamma-ray-induced cascade spectrum at $\gtrsim 30$~TeV energies, which is difficult to observe.  On the other hand, since protons from more distant sources are more depleted, redshift evolution is effectively faster.  Hence, compared to the case of primary gamma rays, contributions from distant sources are more important, and the spectrum is steeper at $\lesssim~$TeV energies for given EBL and redshift evolution models.  The fit also looks better in slower redshift evolution models, especially in negative redshift evolution models that can be expected for, e.g., BL Lac objects in a scenario where Flat Spectrum Radio Quasars (FSRQs) evolve into BL Lac objects at moderate redshifts~\cite{bd02}.  But the required energy budget in no or negative redshift evolution models is found to be inconsistent with the observed CR flux, so we need sufficiently fast evolution models to avoid the overshooting problem~\citep[cf.][]{da11}.  Contrary to the primary gamma-ray injection where the attenuation can be relevant, cascades are essential in the case of the primary CR injection.  The spectral shape is not sensitive to $q$ due to its near-universality, so precise measurements of the VHE DGB can test the VHECR-induced cascade scenario for the VHE DGB.   

\begin{figure*}[bt]
\begin{minipage}{0.49\linewidth}
\begin{center}
\includegraphics[width=\linewidth]{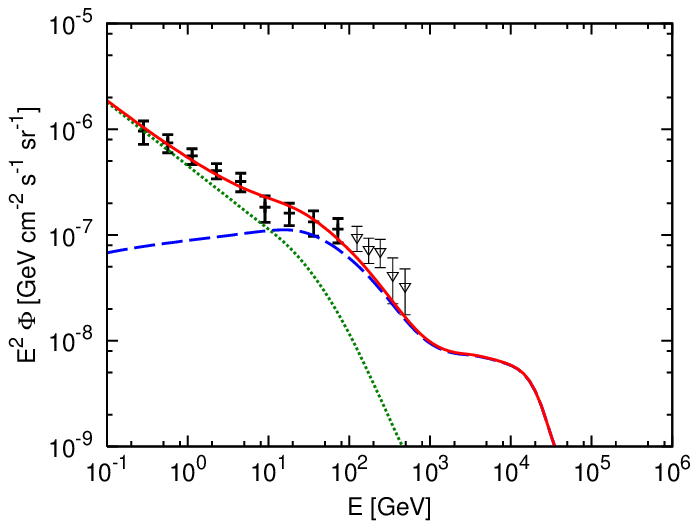}
\caption{
The DGBs from cascades induced by primary VHECRs.  
The proton index is set to $q=2.6$ with $Q_{\rm vhecr}=5.0 \times {10}^{45}~{\rm erg}~{\rm Mpc}^{-3}~{\rm yr}^{-1}$, and no redshift evolution is assumed.  For the power-law component, $a=0.8$ and $\alpha=2.6$ are assumed.  
Note that this case is forbidden by CR observations (see the text).   
}
\end{center}
\end{minipage}
\begin{minipage}{.05\linewidth}
\end{minipage}
\begin{minipage}{0.49\linewidth}
\begin{center}
\includegraphics[width=\linewidth]{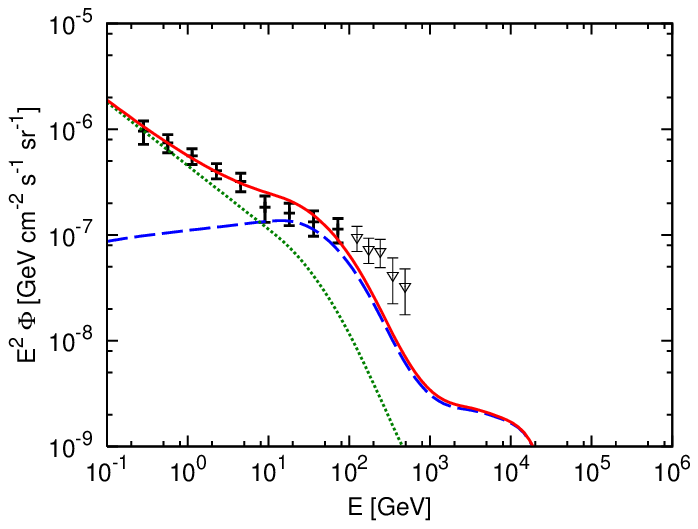}
\caption{The same as Figure~9, but $Q_{\rm vhecr}=7.5 \times {10}^{44}~{\rm erg}~{\rm Mpc}^{-3}~{\rm yr}^{-1}$ with star formation evolution~\cite{hb06}.  
For the power-law component, $a=0.8$ and $\alpha=2.6$ are assumed.  
\newline \, \newline \, \newline}
\end{center}
\end{minipage}
\end{figure*}

The fit with star formation evolution looks to fall below the preliminary \textit{Fermi} data, so the allowed cascade contribution is less than that of the gamma-ray-induced cascade. 
Nevertheless, this VHECR-induced intergalactic scenario seems interesting since it can explain the individual spectra of extreme TeV blazars~\cite{ek10,ess+11,mur+12}, and the cascade hypothesis for the VHE DGB could possibly be consistent with this picture. 
Generally speaking, in order to distinguish between the two possibilities for primary particle injections, it should be relevant to measure the spectrum in the $\sim 0.1-1$~TeV range and at $\gtrsim 30$~TeV energies (cf. Murase et al. 2012~\cite{mur+12}) although it is challenging for current experiments.   

How can we test the cascade hypothesis for the origin of the VHE DGB? 
The anisotropy search would be one of the important probes~\citep[cf.][]{and+07,ack+12}.  
There may be different components from different populations, but even the single population can differently contribute to the DGB because of cascades.  Assuming that the VHE DGB comes from cascades, we may divide the DGB into the secondary component and the primary component.  Then, though the latter may consist of several populations, one typically expects anisotropy change around the transition energy at which cascades become dominant.  The component with cascades is distinguishable from the attenuated primary component only when the void IGMF is not so weak, where resolving more point sources (consisting of primary gamma rays) would in principle lead to a harder DGB around $\sim 100$~GeV.  Such study is left as future work, and we consider a different diagnostic in this work, i.e., implications of TeV surveys with future IACTs such as CTA.

\section{Constraints on energy budgets from extragalactic backgrounds}
In the previous section, we discussed the physics of VHE/UHE gamma rays and showed the potential importance of the cascade components in the VHE DGB, i.e., the DGB at $\gtrsim 100$~GeV, although they can be subdominant as well as other potential contributers. 
In this section, more conservatively, we use the observed DGB flux to constrain the energy budget of gamma rays. 
As seen in the previous section, the spectral shape of cascades is almost universal for VHE/UHE primary particles.  Thus, the DGB flux allows us to access the energy budget of electromagnetic particles with energies higher than $\sim 100$ GeV.  As indicated from Figure~4, the most important energies for setting general cascade bounds correspond to the published data below 100~GeV, so our resulting constraints do not depend on the preliminary \textit{Fermi} data.
Independently, we also restrict the energy budget of high-energy neutrinos from the current background flux upper limit by IceCube and the expected sensitivity of the full IceCube. 
In the case where primary gamma rays are produced through photomeson production, the observed DGB flux can also be useful to constrain the nature of CR sources. 
Analytical estimation of the energy budget of gamma rays and neutrinos is presented in Section 3.1.  In Section 3.2, we numerically obtain constraints on these energy budgets and discuss their implications with particular emphasis on hadronic origin of gamma rays.

\subsection{Analytic estimates}
When one considers the cosmic background flux originating from astrophysical sources, important contribution often comes from sources at $z \sim 1-2$, though details depend on redshift evolution.  
If we take the typical redshift $\bar{z} \sim 1$ to estimate the background flux, because of $E_\gamma^{\rm cut}|_{\bar{z} \sim 1} \sim 0.1$~TeV for cascaded gamma rays, the background gamma-ray flux is crudely written as 
\begin{eqnarray}
E_\gamma^2 \Phi_{\gamma} \approx \frac{c t_H}{4 \pi} (E_\gamma \bar{G}_{E_\gamma} Q_{\gamma}) \xi_z &\simeq& 1.3 \times {10}^{-7}~{\rm GeV}~{\rm cm}^{-2}~{\rm s}^{-1}~{\rm sr}^{-1} \nonumber \\ &\times& \left( \frac{Q_{\gamma}}{6 \times {10}^{44}~{\rm erg}~{\rm Mpc}^{-3}~{\rm yr}^{-1}} \right) \left( \frac{E_\gamma \bar{G}_{E_\gamma}}{0.1} \right) 
\left( \frac{\xi_z}{3} \right),
\end{eqnarray}
where $\bar{G}_{E_{\gamma}}$ is $G_{E_{\gamma}}$ defined in Eq.~(2.2) at $\bar{z} = 1$, $Q_{\gamma}$ is the local bolometric gamma-ray energy budget, and $\xi_z$ is the pre-factor coming from its evolution that is defined as~\cite{wb98}
\begin{equation}
\xi_z \equiv \frac{\int \frac{dz}{1+z} \, \, \left| \frac{dt}{dz} \right| Q_\gamma (z)}{Q_\gamma t_H},
\end{equation}
where $t_H \equiv  \int dz \, \, \left| \frac{dt}{dz} \right|$.  As discussed in the previous section, since cascade gamma rays have a characteristic spectrum, details of the injected gamma-ray spectra do not affect the results much.  In other words, it is not easy to know the intrinsic gamma-ray spectra, and the DGB probes the {\it bolometric} gamma-ray energy input in the VHE/UHE range. 

Contrary to gamma rays, neutrinos can reach the Earth without attenuation in intergalactic space.  In this sense, in principle, neutrino detectors can see the differential energy flux if the number of detected events is enough.  The cumulative neutrino background flux is related to the neutrino energy budget as~\cite{wb98,mb10}   
\begin{eqnarray}
E_\nu^2 \Phi_\nu &\approx& \frac{c t_H}{4 \pi} (E_\nu Q_{E_\nu}) \xi_z \nonumber \\
&\simeq& 3 \times {10}^{-8}~{\rm GeV}~{\rm cm}^{-2}~{\rm s}^{-1}~{\rm sr}^{-1} \left( \frac{E_\nu Q_{E_\nu}}{1.4 \times {10}^{43}~{\rm erg}~{\rm Mpc}^{-3}~{\rm yr}^{-1}} \right) 
\left( \frac{\xi_z}{3} \right),
\end{eqnarray}
where $E_\nu Q_{E_{\nu}}$ is the local differential energy input of neutrinos. 
Although the cumulative neutrino background has not been detected yet, the IceCube-40 integrated limit has now reached $E_\nu^2 \Phi_\nu^{\rm lim} \sim {\rm a~few} \times {10}^{-8}~{\rm GeV}~{\rm cm}^{-2}~{\rm s}^{-1}$~\cite{abb+11}, which gives important constraints on the high-energy budget in neutrinos.

\subsection{Results and implications}
Gamma-ray bounds are obtained by requiring that the resulting spectrum of cascaded gamma rays does not overshoot the total DGB spectrum fitted by Eq.~(2.1) without EBL attenuation.  Here, we calculate the background flux of cascaded gamma rays by using our numerical calculations.  We show the results in Figure~11.  In order to get conservative bounds, we show differential limits assuming the near-mono-energetic injection spectrum for all sources.  They are almost energy-indepedent above $\sim 10$~TeV.  This is because the cascade spectrum is almost universal as expected from Eq.~(2.2), and the electromagnetic energy is essentially conserved between injected gamma rays and reprocessed gamma rays.  
Most of astrophysical sources are likely to have a broader spectrum, and broadening the spectrum around that point brings in additional energy injections.  Hence bounds for power-law injection spectra will be stronger since the DGB is sensitive to the bolometric gamma-ray energy input of VHE/UHE gamma-ray sources.  To demonstrate this, we also show integrated limits assuming $E' Q_{E'}=$const., where tighter bounds are obtained (see Figure~11).  Redshift evolution models moderately affect constraints, and limits are stronger for faster evolution models.  
The DGB obtained by LAT implies that the bolometric gamma-ray energy budget should satisfy
\begin{equation}
Q_{\gamma} = L_\gamma n_s  \lesssim 6 \times {10}^{44}~{\rm erg}~{\rm Mpc}^{-3}~{\rm yr}^{-1} (3/\xi_z),
\end{equation}
for injections in the VHE/UHE range.  

\begin{figure*}[bt]
\begin{minipage}{0.49\linewidth}
\begin{center}
\includegraphics[width=\linewidth]{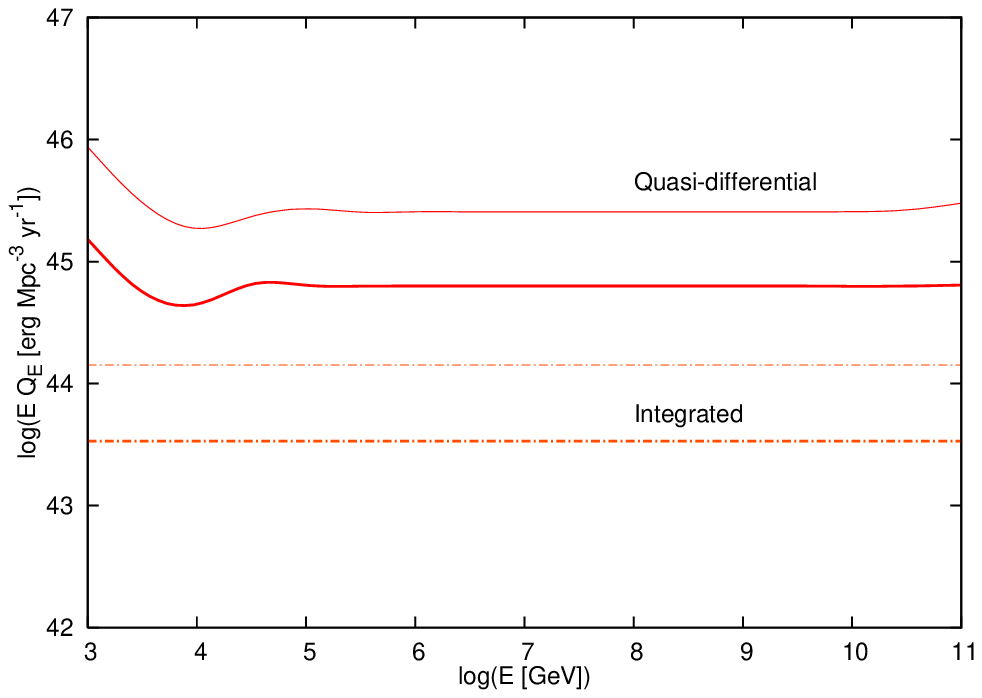}
\caption{Gamma-ray upper limits on the local gamma-ray energy budget of the universe.  
The upper solid curves are quasi-differential limits obtained with $E' Q_{E'} = Q_\gamma$, i.e., the same energy input at any $E'$, for the near-mono-energetic gamma-ray injection. 
The lower dot-dashed curves are integrated limits for $E' Q_{E'} =$~const., i.e., an $E^{-2}$ differential spectrum, where ${E'}_{\rm min}={10}^{2.75}$~GeV and ${E'}_{\rm max}={10}^{11.25}$~GeV are assumed.  The thick curves are obtained for star formation evolution~\cite{hb06}, whereas the thin curves are for no redshift evolution. 
}
\end{center}
\end{minipage}
\begin{minipage}{.05\linewidth}
\end{minipage}
\begin{minipage}{0.49\linewidth}
\begin{center}
\includegraphics[width=\linewidth]{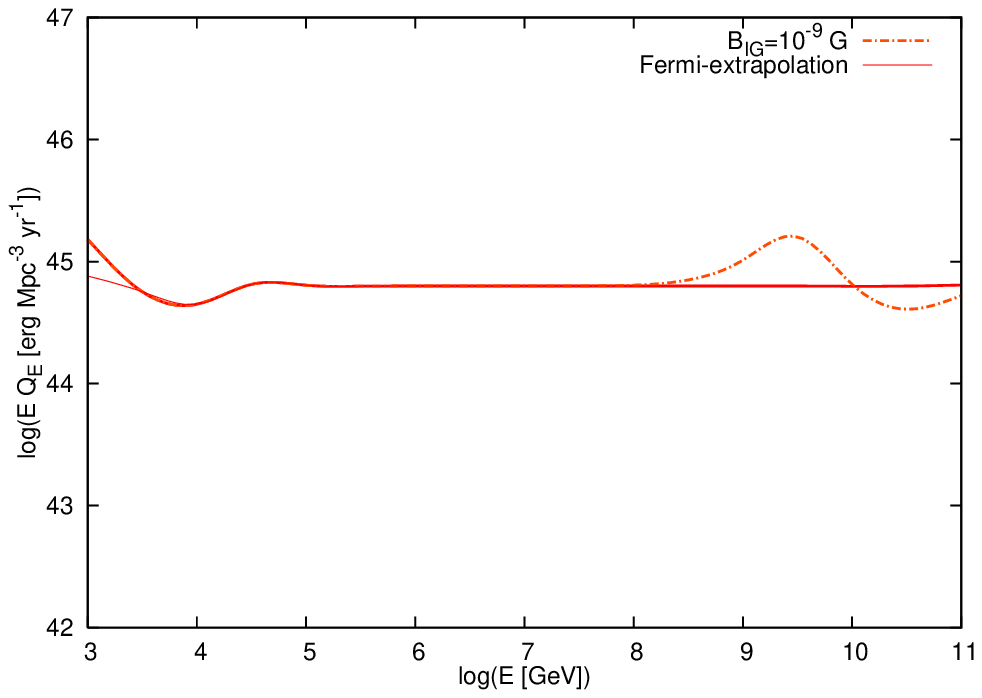}
\caption{The same as Figure~11.  The thick solid curve represents the case where only the DGB fit obtained by Abdo et al.~\cite{abd+10a} is used, while the thin solid curve is for the case where the DGB fit up to $\sim 600$~GeV is used.  In the solid curves, the IGMF is assumed to be negligible for energy losses, while the opposite is assumed for the dot-dashed curve. 
\newline \, \newline \, \newline \, \newline
}
\end{center}
\end{minipage}
\end{figure*}

Cascades gives the bound on the energy budget of VHE/UHE sources as well as a reasonable upper limit on the VHE DGB for some kinds of injections, without specifying sources.  Which kind of astrophysical sources can potentially supply VHE/UHE gamma rays or VHECRs/UHECRs?  Although there are various possibilities, sources detected by \textit{Fermi} may be interesting candidates.  Note that, though their gamma-ray budget in the MeV-GeV range can be estimated from \textit{Fermi} observations, we do not know the VHE/UHE budget of those sources.  This is one of the reasons why the VHE/UHE energy budget constraints are indeed useful. 
Among gamma-ray sources seen by \textit{Fermi}, FR-I galaxies and BL Lac objects have~\cite{dr10} 
\begin{equation}
Q_{\rm [100~MeV,100~GeV]}^{\rm FR-I} \approx {10}^{46}~{\rm erg}~{\rm Mpc}^{-3}~{\rm yr}^{-1},
\end{equation}
whereas FR-II galaxies and FSRQs have 
\begin{equation}
Q_{\rm [100~MeV,100~GeV]}^{\rm FR-II} \sim {10}^{44}-{10}^{45}~{\rm erg}~{\rm Mpc}^{-3}~{\rm yr}^{-1}.
\end{equation} 
Note that the gamma-ray energy budget or the gamma-ray emissivity is expected to be larger at higher redshifts especially for FSRQs.  
Star-burst galaxies, which have recently found in the VHE range for nearby sources~\cite{ace+09,acc+09}, may also supply the comparable energy budget,
\begin{equation}
Q_{\rm [100~MeV,100~GeV]}^{\rm SG} \sim {10}^{46}~{\rm erg}~{\rm Mpc}^{-3}~{\rm yr}^{-1},
\end{equation}
and normal starforming galaxies may also be relevant. 
GRBs are the most violent explosions in the universe, but their gamma-ray energy budget is expected to be smaller in the LAT band than guaranteed DGB contributers~\citep[see, e.g.,][]{ste+93,pf02,lw06,tho+06,der07,kne07}, and only a fraction of GRBs have been detected by LAT.  Most of the electromagnetic energy seems to be radiated in the MeV range, where one expects~\cite{lia+07,ld07,wp10}
\begin{equation}
Q_{\rm [keV,10~MeV]}^{\rm GRB} \approx (0.3-2) \times {10}^{44}~{\rm erg}~{\rm Mpc}^{-3}~{\rm yr}^{-1},
\end{equation}
and that resulting intergalactic cascade radiation is studied in detail by Refs.~\cite{cas+07,mur+07}. 

From Eqs.~(3.5) and (3.7), the VHE/UHE budget of BL Lac objects and star-burst galaxies in VHE/UHE gamma rays or VHECRs/UHECRs is expected to be smaller than $\sim 5-30$~\% of their gamma-ray energy budget in the $\sim 100$~MeV range.  For BL Lac objects, since the GeV gamma-ray emission is typically attributed to leptonic processes, cascaded gamma rays that can contribute to the VHE DGB could originally come from the same process but a sub-class, or by different processes including CR-induced processes.  For star-burst galaxies, the GeV emission is believed to be hadronic and the spectrum is is typically steeper than $s=2$, so it is not natural to invoke excessive VHE/UHE gamma-ray emissions.  For FSRQs and GRBs, values suggested in Eqs.~(3.6) and (3.8) are comparable or smaller than that from the DGB constraint.  Especially for GRBs, the constraint implies that the missing energy budget in the VHE/UHE range has to be quite large in order to make a significant part of the VHE DGB via cascades.  
Structure formation shocks, though gamma rays have not been detected from them, are also interesting sources of the DGB~\cite{lw00,kes+03}.  The kinetic energy budget is order of $\sim {10}^{48}~{\rm erg}~{\rm Mpc}^{-3}~{\rm yr}^{-1}$~\citep[e.g.,][]{mur+08a,ski+08}, so the DGB constraint implies that only a fraction of the energy can be used for gamma rays (and neutrinos). 
Each class of known astrophysical sources has typical values of the luminosity $L$ and source number density $n_s$.  Therefore, not only the DGB constraint but also the constraint placed by searches for individual sources should also be useful.  In Section~4, we also show the importance and power of the IACT survey constraint.  

The energy budget constraint is very important to get implications for various kinds of more exotic physics.  It enables us to put interesting constraints on the properties of annihilating and decaying dark matter~\citep[e.g.,][and refrences therein]{mb12}.  Also, it can be used for constraining the abundance of primordial black holes that could contribute to the DGB via Hawking radiation~\cite{car+10}.  

Note that the DGB constraints are essentially determined by the DGB around $\sim 100$~GeV.  In fact, the results are not changed much if Eq.~(2.1) is extrapolated up to $\sim 600$~GeV (see Figure~12).  The IGMF can moderately affect the results only in the UHE range.  In Figure~11, the IGMF is neglected in calculating the cascaded gamma-ray flux, but it may be strong enough that the synchrotron cooling is more important than the IC cooling at UHE.  In Figure~12, we show the case for an {\it effective} IGMF of $B_{\rm IG}=1$~nG, which is consistent with upper limits by Faraday rotation measurements~\citep[e.g.,][and references therein]{ko11,tm12}~\footnote{IGMFs in the universe are likely to be inhomogeneous, that is, the void IGMF can be much weaker whereas structured extragalactic magnetic fields can be much stronger.}  
From Figure~12, one sees that the results can be affected by a factor of $\sim 2$ due to the IGMF only at energies $\gtrsim {10}^{8}$~GeV.   
 
The obtained gamma-ray limits could be improved by resolving more point sources.  Several authors have recently attempted to estimate contributions of the point sources below the LAT detection limit~\cite{abd+10c,mh11,aba+11,aba+12}, and the contribution of blazars is typically estimated to be $\sim 10-20$~\%.  We will have better constraints on the gamma-ray energy budget if more point sources contributing to the DGB are resolved.  In this work, we obtained gamma-ray limits based on the full DGB, which is conservative since the origin of the DGB is not yet agreed upon. 

\begin{figure*}[bt]
\begin{minipage}{0.49\linewidth}
\begin{center}
\includegraphics[width=\linewidth]{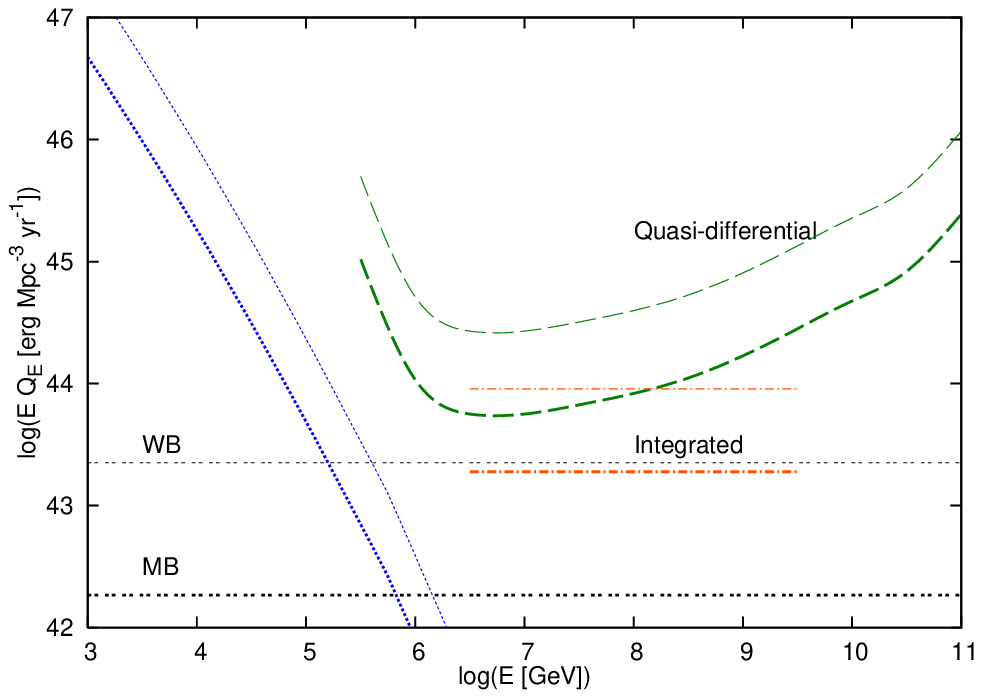}
\caption{Neutrino bounds on the local neutrino energy budget of the universe.  The upper dashed curves represent quasi-differential limits estimated from the IceCube-40 analysis in 333.5~days~\cite{abb+11}.  The dotted curves  show limits from the atmospheric neutrino background.  The lower dot-dashed curves represent integrated limits obtained for $E_{\nu} Q_{E_{\nu}} =$~const.  The thick curves are obtained for star formation evolution~\cite{hb06}, whereas the thin curves are for no redshift evolution.  WB represents the Waxman-Bahcall bound~\cite{wb98}, whereas MB is for the Murase-Beacom bound for the effective iron survival~\cite{mb10}.
}
\end{center}
\end{minipage}
\begin{minipage}{.05\linewidth}
\end{minipage}
\begin{minipage}{0.49\linewidth}
\begin{center}
\includegraphics[width=\linewidth]{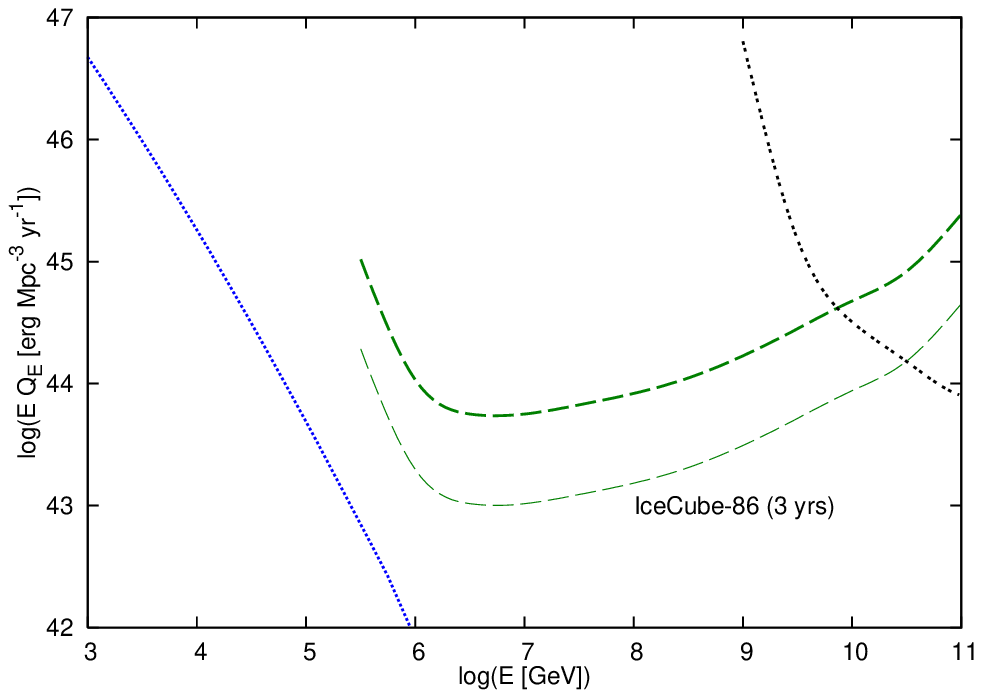}
\caption{The same as Figure~13.  The thick dashed curve is for the IceCube-40 333.5~day observation, whereas the thin dashed curve is for the future full IceCube 3~year observation~\cite{abb+11}.  The ANITA limit is shown with the double-dotted curve~\cite{gor+10}.  The atmospheric neutrino background limit is overlaid with the dotted line~\cite{bec08}.  Star formation evolution is assumed. 
\newline \, \newline \, \newline \, \newline \, \newline
}
\end{center}
\end{minipage}
\end{figure*}

Next, let us consider constraints that can be placed by high-energy neutrino detectors such as IceCube.  In order to obtain constraints on the neutrino energy budget, we have used a simplified approach through Eq.~(3.3), and we get
\begin{equation}
E_{\nu} Q_{E_{\nu}} \lesssim 1.4 \times {10}^{43}~{\rm erg}~{\rm Mpc}^{-3}~{\rm yr}^{-1}~\left( \frac{E_\nu^2 \Phi_\nu^{\rm lim}}{3 \times {10}^{-8}~{\rm GeV}~{\rm cm}^{-2}~{\rm s}^{-1}~{\rm sr}^{-1}}\right) (3/\xi_z).  
\end{equation}
We show the results obtained from the IceCube-40 data~\cite{abb+11} in Figure~13, which suggests that the neutrino constraints can be even stronger than the gamma-ray constraints for hadronic sources that gamma rays and neutrinos are comparably produced.  As seen in Figure~14, they should be improved by full IceCube observations, where the expected sensitivity is assumed~\cite{abb+11}.  
Note that the neutrino limit becomes worse at lower energies, due to atmospheric neutrinos.  
Generally speaking, it depends on details of sources (e.g., gamma-ray/neutrino production channels, a spectral index for CRs, and so on) which is stronger, the gamma-ray constraint or the neutrino constraint.  But Figure~13 shows that neutrino observations are powerful as a probe of the VHE/UHE universe.   

For astrophysical neutrino emitters, i.e., CR accelerators, high-energy neutrinos are produced via hadronuclear and/or photonuclear reactions.  In the case of photomeson production that occurs when an incident photon energy (in the ion rest frame) exceeds the pion mass, since $\pi^{+}:\pi^0 \sim 1:1$ and a neutrino carries a quarter of the pion energy, the cumulative neutrino background is estimated to be~\citep[e.g.,][]{wb97,mb10}
\begin{equation}
E_\nu Q_{E_\nu} \approx \frac{3}{8} {\rm min}[1,f_{\rm mes}] (E_{\rm cr} Q_{E_{\rm cr}}),
\end{equation}
where $f_{\rm mes}=f_{\rm mes}(E_{\rm cr})$ is the photomeson production efficiency that depends on sources.  For a power-law target photon spectrum, $dn/d\varepsilon = n_0 {(\varepsilon/\varepsilon_0)}^{-\beta}$ ($\beta \gtrsim 1$), we have 
\begin{equation}
f_{\rm mes} \approx \frac{2 \varepsilon_0 n_0}{1+\beta} c t_{\rm int} \sigma_{\Delta} \frac{{\Delta \bar{\varepsilon}}_\Delta}{\bar{\varepsilon}_\Delta} {\left( \frac{E_{\rm cr}}{E_{\rm cr}^0} \right)}^{\beta-1},   
\end{equation}
where $t_{\rm int}$ is the interaction time between CRs and photons, $\sigma_\Delta \sim 5 \times {10}^{-28}~{\rm cm}^2$, $\kappa_\Delta \sim 0.2$, $\bar{\varepsilon}_\Delta \sim 0.3$~GeV,
$\Delta \bar{\varepsilon}_\Delta \sim 0.2$~GeV, $E_{\rm cr}^0 \approx 0.5 \delta^2 \bar{\varepsilon}_{\Delta} m_p c^2/\varepsilon_0$, and $\delta$ is the Doppler factor of the CR source.  For example, in the case of GRB prompt emission, one typically expects $f_{\rm mes} \sim 0.01-0.1$~\cite{wb97}.

The formal limit of semi-transparent sources, $f_{\rm mes} \rightarrow 1$, is often referred as the Waxman-Bahcall bound~\cite{wb98}, which implies
\begin{equation}
E_\nu Q_{E_\nu} \lesssim 2.3 \times {10}^{43}~{\rm erg}~{\rm Mpc}^{-3}~{\rm yr}^{-1} \left( \frac{E_{\rm cr} Q_{E_{\rm cr}}}{0.6 \times {10}^{44}~{\rm erg}~{\rm Mpc}^{-3}~{\rm yr}^{-1}} \right) 
\end{equation}
for CRs with $E_{\rm cr} Q_{\rm cr}={\rm const.}$, where $E_{\rm cr} Q_{E_{\rm cr}} \approx 0.6 \times {10}^{44}~{\rm erg}~{\rm Mpc}^{-3}~{\rm yr}^{-1}$ is the UHECR energy budget that depends on their spectral index~\cite{kat+09}.  Though more general neutrino bounds are obtained in Ref.~\cite{man+01}, many of the theoretical predictions on neutrinos from UHECR sources lie below those bounds.  Also, as claimed by the IceCube collaboration~\cite{abb+11} and indicated in Figure~13, the current IceCube-40 results on the neutrino background have already reached this bound for fast redshift evolution.  
The Waxman-Bahcall bound, i.e., effective nucleon-survival bound is shown in Figure~13 with $E_{\rm cr} Q_{E_{\rm cr}} =0.6 \times {10}^{44}~{\rm erg}~{\rm Mpc}^{-3}~{\rm yr}^{-1}$.

If observed UHECRs consist of heavy nuclei rather than protons, CRs have to survive photodisintegration both inside and outside the source.  Requiring survival of nuclei implies that the number of target photons necessary for neutrino production is also limited, and Murase \& Beacom~\cite{mb10} gives  
\begin{equation}
E_\nu Q_{E_\nu} \lesssim 1.8 \times {10}^{42}~{\rm erg}~{\rm Mpc}^{-3}~{\rm yr}^{-1}~{\left( \frac{A}{56} \right)}^{-0.21} \left( \frac{E_{\rm cr} Q_{E_{\rm cr}}}{0.6 \times {10}^{44}~{\rm erg}~{\rm Mpc}^{-3}~{\rm yr}^{-1}} \right),
\end{equation}
for CRs with $E_{\rm cr} Q_{\rm cr}={\rm const.}$, which is order of magnitude smaller than Eq.~(3.12).  It is also applicable to cosmogenic neutrinos when UHECRs arriving \textit{at the earth} are still nucleus-rich, and it is indeed consistent with recent detailed calculations~\citep[see][and references therein]{anc+07,ko11}.  As indicated in Figure~13, the effective iron-survival bound is comparable to the integrated limit placed by the full IceCube in three years~\cite{mb10}.

CR accelerators should produce gamma rays as well as neutrinos.  However, regarding \textit{primary} gamma rays and neutrinos in the sense that they are produced as secondaries by CR ions or electrons \textit{inside} the sources, the amount of VHE/UHE gamma rays inducing the intergalactic cascade highly depend on details of source physics.
If hadronic gamma rays are produced by photonuclear reactions (rather than ion synchrotron radiation) and the source is optically thin for the pair creation process, in the case of proton accelerators, there are three contributions.
The first and second contributions are $\pi^0$ gamma rays and electron-positron pairs from photomeson production.  The third contribution consists of pairs from the Bethe-Heitler process.  
If the photomeson production is dominant, neutrino constraints by IceCube are stronger than gamma-ray constraints at sufficiently high energies~\citep[see also][]{ahl+10}.
If the Bethe-Heitler process is more relevant than the photomeson production in the total proton energy loss, gamma-ray constraints can be more stringent than neutrino constraints.  

However, for powerful synchrotron sources such as GRBs and AGN, charged leptons are confined and VHE/UHE gamma rays are cascaded inside the source, and a significant fraction of the electromagnetic energy may be radiated as lower-energy gamma rays below even the LAT band. When the source cascade is relevant, Eq.~(2.2) no longer applies and the DGB constraint can be irrelevant depending on the pair creation opacity in the source. 
On the other hand, a fraction of gamma rays can escape from the source and can induce intergalactic cascades~\cite{mur09}, where the DGB constraint is applicable.  
For such sources, the released gamma-ray energy flux is comparable to the released neutrino energy flux at energies where escape is possible, and the neutrino constraints shown in Figures~13 and 14 are stronger than the gamma-ray constraint shown in Figures~11 and 12.  
In Figure~8, we showed the case where UHE gamma rays contribute to the DGB.  However, as long as neutrinos are expected (though neutrino detection at $\gtrsim$~EeV is more difficult than at $\sim$~PeV), it is unlikely that UHE gamma rays contribute to the DGB.     

Finally, we comment on the case of the intergalactic VHECR-induced cascade.  The situation is similar but more or less different for secondaries produced {\it outside} the source, i.e., gamma rays and neutrinos produced via interactions of CRs leaving the source with the EBL and CMB.  In this case, the Bethe-Heitler process can be dominant especially when the spectrum of injected CRs is steep and/or their maximum energy is not high enough to cause photomeson production.  Detailed comparisons between the gamma-ray constraint and the neutrino constraint have been recently done in the specific context of UHECRs~\cite{ahl+10,ber+11,da11,gel+11,wan+11}, so we do not discuss them further.

\section{Constraints from TeV surveys of individual sources}
\subsection{Analytical estimates}
Constraints from the diffuse or cumulative backgrounds are valid for a general class of extragalactic gamma-ray sources.  From Eq.~(3.4), the DGB measured by LAT gives a constraint of
\begin{equation}
n_s \lesssim 2 \times {10}^{-6}~{\rm Mpc}^{-3}~{\left( \frac{L_\gamma}{{10}^{43}~{\rm erg}~{\rm s}^{-1}}  \right)}^{-1}~{\left( \frac{f_z}{3}\right)}^{-1},  
\end{equation}
as long as the injected gamma-ray energy is high enough. Note that the above expression holds in the CR-induced cascade case.  When the cascade is governed by the Bethe-Heitler process, we expect that the main contribution comes from distant VHECR sources. Since VHE/UHE protons are expected to be depleted during their propagation over cosmological distance, one expects 
\begin{equation}
n_s \lesssim 2 \times {10}^{-6}~{\rm Mpc}^{-3}~{\left( \frac{\bar{f}_{\rm BH} L_{\rm vhecr}}{{10}^{43}~{\rm erg}~{\rm s}^{-1}}  \right)}^{-1}~{\left( \frac{f_z}{3}\right)}^{-1},  
\end{equation}
where $L_{\rm vhecr}$ is the total luminosity of VHECRs that can efficiently interact with CMB photons via the Bethe-Heitler process, $\bar{f}_{\rm BH}$ is the typical Bethe-Heitler efficiency.  Note that this constraint becomes weaker as the proton maximum energy is lower. 

Not only measurements of the diffuse or cumulative backgrounds but also surveys of individual (point and extended) sources can give us useful information on the population of luminous steady sources~\citep[e.g.,][]{aha+06}.  
The cascade can make sources more detectable than the case where the attenuation alone is taken into account.
Searches for individual sources would be especially important for IACTs such as CTA, since it is difficult for them to detect the DGB in the TeV range because of their small fields of view~\citep[see, e.g.,][for the study on blazars]{ino+10}.  
In the following, we discuss the detectability of an individual source through surveys with next-generation IACTs such as CTA, in addition to background constraints by LAT and IceCube.
In particular, we consider survey constraints in the $n_s-L$ plane, assuming non-detections in the total observation time $t_{\rm sur}$.  For our purposes, we take the DGB constraint and IACT survey constraint to be independent and complementary.  

By the time when CTA starts observations, the DGB measured by \textit{Fermi} may be reduced by resolving more point sources, and the DGB constraint could be stronger.  But we conservatively use the current DGB constraint in this work.   
If the IACT survey constraint is weaker than the DGB constraint, sources that contribute to the DGB via cascades will be allowed to make a significant part of the VHE DGB.  If the IACT survey constraint becomes stronger, which is expected especially for luminous VHE sources, it can exclude scenarios where those luminous sources make an important contribution to the VHE DGB.  
This implies that constraints by IACTs on populations of undetected sources can be more powerful than what is inferred from the \textit{Fermi} DGB measurement that enables us to probe only the overall energy budget.  The joint-analyses would be helpful especially when the constraints are comparable.  

In Section~2, we used $Q(z)$, since the DGB depends only on the energy budget. But, in following discussions, we need to break the degeneracy between and $L (z)$ and $n(z)$.   
Hereafter, for simplicity, we assume the limit that $Q (z) = L n_s (z)$, i.e., the luminosity does not have redshift evolution.  This may not be the case in, e.g., AGN, where the pure luminosity evolution has often been considered~\citep[but see, e.g.,][]{it09,aba+11}.  
Studying this case is conservative, since the other limit, $Q (z) = L(z) n_s$, allows us to see more distant sources (see Eq.~(4.7)) for the same redshift evolution, which is more optimistic.  Also, we do not consider the luminosity function since we do not specify any class of gamma-ray/neutrino or VHECR sources.  

At the Earth, the energy flux of a source with the isotropic luminosity $L$ is written as
\begin{equation}
E F_E = \frac{1}{4 \pi d^2} (E {\mathcal G}_E L)
\end{equation}
where $E {\mathcal G}_E$ is introduced as 
\begin{equation}
E {\mathcal G}_E \equiv
\left\{ \begin{array}{ll}
(E L_E/L)
& \mbox{(neither attenuation nor cascade)}\\
(E L_E/L) e^{-\tau}
& \mbox{(attenuation)}\\
(E G_E)  
& \mbox{(cascade)},
\end{array} \right. 
\end{equation}
where $G_E$ is introduced in Eq.~(2.2).  
For neutrinos, we consider neither attenuation nor cascade.  For gamma rays, we take into account either attenuation or cascade.  
For primary gamma-ray injections, the attenuation case is conservative and more realistic when the secondary contribution is negligible.  For collimated sources like blazars, it is the case when the void IGMF is so strong that pairs are deflected and may even be isotropized (and an giant isotropic pair halo is formed)~\cite{aha+94}.  On the contrary, the cascade cases are justified in such collimated sources when the IGMFs are weak enough.  We especially assume that cascaded TeV gamma rays are seen as point sources~\footnote{Weak IGMFs can still be important for transient collimated sources, though this work focuses on steady sources.  Noticing that $E_\gamma^{\rm IC} \approx (4/3) \gamma_e^2 \varepsilon_{\rm CMB}$, the corresponding time spread at TeV energies is~\cite{mur+08c} is ${\Delta t}_{\rm IG} \approx (1/2) \theta_{\rm IG}^2 (\lambda_{\gamma \gamma}/c) \simeq 430~{\rm yr}~B_{\rm IG,-15}^2 (\lambda_{\gamma \gamma}/300~{\rm Mpc}) {(E_{\gamma}/\rm TeV)}^{-2}$, which can be longer than the source duration.}, and this assumption is independently motivated by cascade interpretations of extreme TeV blazars.  In the intermediate regime, secondary GeV-TeV gamma rays form an anisotropic pair halo around a collimated source~\cite{ns07,mur+08c}.  The deflection angle of electron-positron pairs is estimated to be $\theta_{\rm IG} \approx (\sqrt{2} \lambda_{\rm IC})/(\sqrt{3} r_L) \simeq 0.063~{\rm deg}~B_{\rm IG,-15} \gamma_{e,7.5}^{-2}$, so the anisotropic pair halo becomes relevant when $\theta_{\rm IG} > \theta_j$ (that is the jet opening angle) for collimated sources such as blazars and GRBs.  In the case of primary gamma-ray injections, its apparent size is estimated to be $\Theta_{\rm ex} \approx (\lambda_{\gamma \gamma}/d) (1+\lambda_{\gamma \gamma}/d) \theta_{\rm IG} \simeq 0.02~{\rm deg}~(\lambda_{\gamma \gamma}/300~{\rm Mpc}) ({\rm Gpc}/d) B_{\rm IG,-15} \gamma_{e,7.5}^{-2} (1+\lambda_{\gamma \gamma}/d)$.  If the angular resolution of $\sim 0.05$~deg is achieved at TeV~\cite{act+11}, the source extension is irrelevant only when the void IGMF is weaker than a few~$\times {10}^{-15}$~G for distant sources.  For isotropic sources, we expect $\Theta_{\rm ex} \approx (\lambda_{\gamma \gamma}/d) \simeq 0.3~{\rm deg}~(\lambda_{\gamma \gamma}/300~{\rm Mpc}) ({\rm Gpc}/d)$.  

Note that ${\mathcal G}_E$ also depends on $d$.  Then, the flux is compared to detector sensitivities.  In this work, as the detection criterion, we use 
\begin{equation}
F_E \geq F_E^{\rm lim}, 
\end{equation}
where $F_E^{\rm lim}$ is the detector sensitivity in unit of particles per time per area.  
We use the differential sensitivity in this study, though the integrated one would improve the detectability by a factor.  Measurements of spectra are eventually necessary to reveal source physics in detail, and the differential sensitivity can be easily compared to theoretical predictions~\cite{ino+10}.  
Assuming a survey with its total observational area ${\Delta \Omega}_{\rm sur}$ and total observation time $t_{\rm sur}$, the individual source sensitivity becomes  
\begin{equation}
F_E^{\rm lim} = F_E^{\rm sur} {\left( \frac{t_{\rm sur}}{t_{\rm fov}} \right)}^{1/2}
\end{equation}
where $t_{\rm fov} = t_{\rm sur} ({\Delta \Omega}_{\rm fov}/{\Delta \Omega}_{\rm sur})$ is the observation time per patch and we set the field of view to ${\Delta \Omega}_{\rm fov}=20~{\rm deg}^2$~\cite{aha+06}.  
In this study, we focus on next-generation IACTs, and we use the CTA differential sensitivity for the 50~hr observation, $F_E^{50}$~\cite{act+11}.  We also fix the total survey time, i.e., how much time is devoted to pointing observations, to $t_{\rm sur}=1200$~hr.  This is possible by multi-year observations, and $t_{\rm sur}=250$~hr is comparable to the time dedicated by HESS to its Galactic plane survey and it will represent only 1/4 of the observation time expected for CTA during the first year of observations.  Of course, the resulting constraints can be improved if a longer survey time is used. 


For given $L$, one can define the distance above which we cannot detect a single source as 
\begin{equation}
d_{\rm lim} \equiv {\left( \frac{E {\mathcal G}_E L}{4 \pi E F_E^{\rm lim}} \right)}^{1/2}.
\end{equation}
The number of sources located in a cone with a radius $d_{\rm lim}$ and a solid angle ${\Delta \Omega}_{\rm sur}$ is $n_s  ({\Delta \Omega}_{\rm sur}/3) d_{\rm lim}^3$, so the critical number density of sources, where the number within the cone becomes unity, is~\citep[see, e.g.,][]{lip06,sb10}  
\begin{equation}
n_c = \left( \frac{3}{{\Delta \Omega}_{\rm sur} d_{\rm lim}^3} \right) = \frac{3}{{\Delta \Omega}_{\rm sur}} {\left( \frac{4 \pi E F_E^{\rm lim}}{E {\mathcal G}_E L} \right)}^{3/2} 
\end{equation}
Then, using Eq.~(4.5), a non-detection with a CTA-like IACT survey would imply 
\begin{eqnarray}
n_s \lesssim n_c &\sim& 6 \times {10}^{-7}~{\rm Mpc}^{-3}~{\left( \frac{L}{{10}^{43}~{\rm erg}~{\rm s}^{-1}}  \right)}^{-3/2} {\left( \frac{E {\mathcal G}_{E}}{0.2}  \right)}^{-3/2}  {\left( \frac{t_{\rm sur}}{\rm 250~hr} \right)}^{-3/4} \nonumber \\ 
&\times& {\left( \frac{E F_E^{50}}{{10}^{-13}~{\rm erg}~{\rm cm}^{-2}~{\rm s}^{-1}}  \right)}^{3/2}  {\left( \frac{{\Delta \Omega}_{\rm sur}}{4000~{\rm deg}^2}  \right)}^{-1/4} 
{\left( \frac{{\Delta \Omega}_{\rm fov}}{{20}~{\rm deg}^2}  \right)}^{-3/4},
\end{eqnarray}
where we have used $E F_E^{\rm sur} \propto t_{\rm sur}^{-1/2}$. 
Comparing Eq.~(4.9) to Eq.~(4.1), one sees that the DGB constraint is more stringent for less-luminous gamma-ray sources and vice versa for more-luminous gamma-ray sources.   
For neutrino observations by IceCube, the background limit is typically stronger than the point source limit except for nearby, special (exceptional) sources~\cite{lip06,sb10}. 

One can make a similar argument for the VHECR-induced cascade. For the VHECR-induced cascade, $E G_E$ follows 
\begin{equation}
E G_E \propto \frac{\tau_{\rm BH}}{\tau_{\gamma \gamma}} (1-e^{-\tau_{\gamma \gamma}}) \sim 
\left\{ \begin{array}{ll}
\frac{d}{\lambda_{\rm BH}}
& \mbox{($d < \lambda_{\gamma \gamma}$)}\\
\frac{\lambda_{\gamma \gamma}}{\lambda_{\rm BH}}
& \mbox{($\lambda_{\gamma \gamma} < d < \lambda_{\rm BH}$)}.
\end{array} \right. 
\end{equation}  
Therefore, the luminosity-dependence of $n_s$ can be different from Eq.~(4.9). 
When $\lambda_{\gamma \gamma}$ at the typical energy of observed photons is longer than $d_{\rm lim}$, we expect  
\begin{equation}
n_c \propto L_{\rm vhecr}^{-3} t_{\rm sur}^{-3/2} {(E F_E^{50})}^{3} {\Delta \Omega}_{\rm sur}^{1/2} 
{\Delta \Omega}_{\rm fov}^{-3/2},
\end{equation}
which is different from the conventional relation
\begin{equation}
n_c \propto L_{\rm vhecr}^{-3/2} t_{\rm sur}^{-3/4} {(E F_E^{50})}^{3/2} {\Delta \Omega}_{\rm sur}^{-1/4} 
{\Delta \Omega}_{\rm fov}^{-3/4}.
\end{equation} 
The reason why Eq.~(4.11) is different from Eq.~(4.12) is that more gamma rays are produced during the CR propagation for more distant CR sources, since the CR energy loss due to the Bethe-Heitler process increases as $d$ unless CRs are depleted.

\subsection{Results and prospects}
In this subsection, we show the numerical results on constraints in the $n_s-L$ plane that come from the DGB measured by LAT, TeV surveys by next-generation IACTs, and the neutrino background limit by IceCube.  Taking into account cosmology, the source number density is numerically calculated by
\begin{equation}
n_c = {\left[ \frac{c}{H_0} \int^{z_{\rm lim}} dz \,\, \frac{\Delta \Omega_{\rm sur} d_L^2 (n_s(z)/n_s)}{{(1+z)}^2 \sqrt{\Omega_\Lambda+{(1+z)}^3 \Omega_m}}  \right]}^{-1},
\end{equation}
where $d_L$ is the luminosity distance and $z_{\rm lim}$ is the redshift corresponding to $d_{L}^{\rm lim}$.  Throughout this section, for demonstrative purposes, we take the star formation history as a redshift evolution model~\cite{hb06}. 

\begin{figure*}[bt]
\begin{minipage}{0.49\linewidth}
\begin{center}
\includegraphics[width=\linewidth]{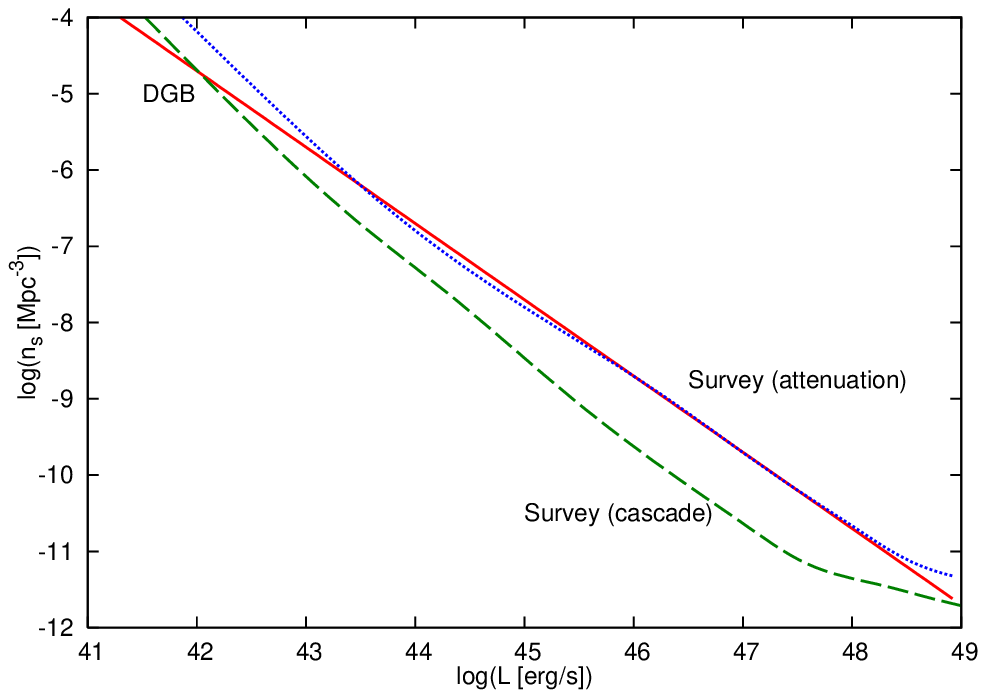}
\caption{Constraints from the DGB measured by \textit{Fermi} and from TeV surveys with future CTA-like IACTs.  
Here $s=1.5$ and ${E'}_{\rm max}={10}^{2.25}$~TeV is assumed for the primary gamma-ray spectrum.  Whereas Survey (attenuation) represents the case where only the attenuated emission is relevant, Survey (cascade) represents the case where the cascade emission contributes to the individual source flux.}
\end{center}
\end{minipage}
\begin{minipage}{.05\linewidth}
\end{minipage}
\begin{minipage}{0.49\linewidth}
\begin{center}
\includegraphics[width=\linewidth]{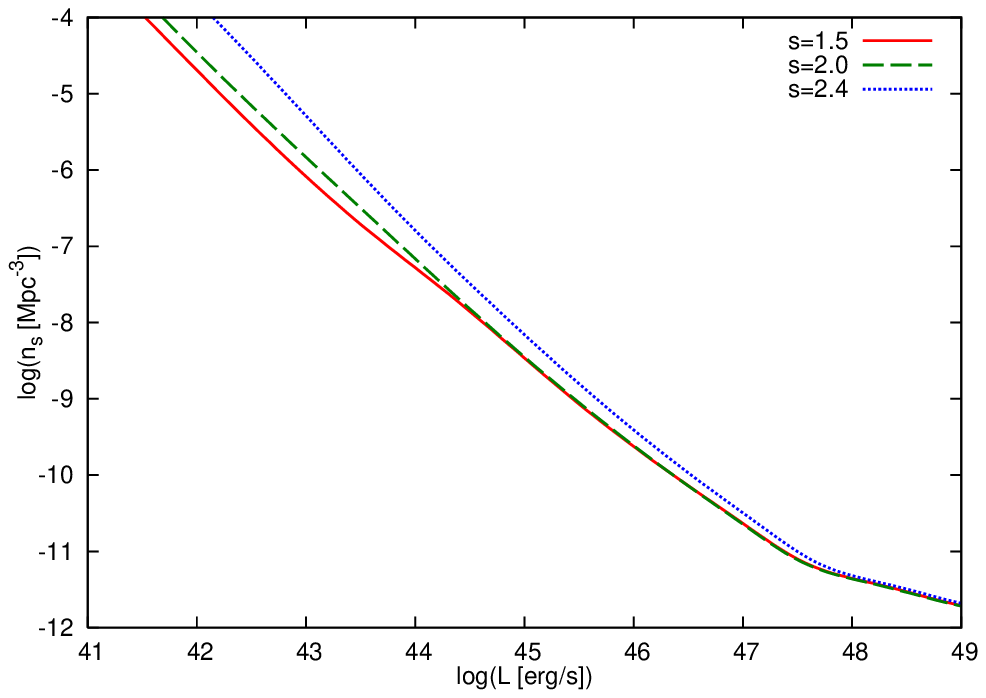}
\caption{Dependence on photon indices for the injection spectrum, in constraints from TeV surveys with future CTA-like IACTs.  The primary gamma-ray spectrum is normalized with ${E'}_{\rm max}={10}^{2.25}$~TeV for $s=1.5$, ${E'}_{\rm min}={10}^{-2.25}$~TeV and ${E'}_{\rm max}={10}^{2.25}$~TeV for $s=2.0$, and ${E'}_{\rm min}={10}^{-2.25}$~TeV for $s=2.4$, respectively.  Cascade contributions are included. 
\newline}
\end{center}
\end{minipage}
\end{figure*}

First, we consider VHE gamma-ray emitters with hard injection spectra.  Such sources have been found in extreme TeV blazars by IACT observations~\cite{aha+07,aha+07b}, as mentioned above.  The results for $E_\gamma^\prime L_{E_\gamma^\prime} \propto {(E_{\gamma}^\prime)}^{2-s}$ with $s=1.5$ are shown in Figure~15.  
In the attenuation case, the constraint from the DGB is almost comparable or more stringent at low luminosities, so it may be difficult for IACT surveys to find them.  However, in the cascade case, the future IACT survey constraint is stronger than the DGB constraint for luminous sources with $L_\gamma \gtrsim {10}^{42}~{\rm erg}~{\rm s}^{-1}$.  Here, one should keep in mind that current IACT limits seem to be weaker and worse than the DGB constraint, since their sensitivities are $\sim 10$ times lower than the CTA sensitivity. 
Note that the dependence expected from the analytical estimate, $n_c \propto L_\gamma^{-1.5}$, can be seen only at sufficiently low luminosities, and deviation from this analytical relation is more important at higher luminosities.  This is because higher luminosities correspond to larger values of $d_L^{\rm lim}$ and the EBL attenuation becomes more significant at larger distances.    
The dependence on $s$ is shown in Figure~16.  Due to the near-universal cascade spectrum, the result is not sensitive once the spectral index is harder than 2 as long as ${E'}_{\gamma}^{\rm max}$ is high enough.  Obviously, LAT observations in the GeV range are more relevant for steeper ($s>2$) photon indices.  

In Figure~17, we consider UHE gamma rays~\cite{mur09,mur12}, which can be produced in UHECR accelerators.  
Again, the qualitative results do not change compared to those shown in Figure~15, since cascades lead to a near-universal spectrum for distant sources.  This is seen from the fact that the result is not much changed even for the primary gamma-ray spectrum with $E'_\gamma Q_{E'_\gamma}=$~const.  Note that the constraint becomes significantly loosened at low luminosities since a significant fraction of UHE gamma rays from nearby sources can reach the Earth~\cite{mur09}. 

Next, we consider the VHECR-induced intergalactic cascade, which provides an interesting possibility to explain extreme TeV blazars~\cite{ess+10,ess+11,mur+12,raz+12}.  This case is shown in Figure~18, but we see qualitative changes compared to those shown in Figures~15 and 17.  
At lower luminosities, Eq.~(4.11) rather than Eq.~(4.12) is more appropriate since gamma rays are mainly injected via the Bethe-Heitler process.  At high luminosities, the dependences become more or less similar.  But, due to the Bethe-Heitler process, the VHECR-induced cascade is easier to see for distant sources, at higher energies such that $\lambda_{\gamma \gamma} < d$ (as long as the IGMFs are weak enough)~\citep[see also Figure~7 of Ref.][]{mur+12}.  
Note that, in the VHECR-induced cascade scenario of extreme blazars, the apparent isotropic luminosity of $L_{\rm vhecr} \sim {10}^{45-46}~{\rm erg}~{\rm s}^{-1}$ is typically required, depending on structured extragalactic magnetic fields~\cite{mur+12,raz+12}.  For such luminosities, the IACT survey constraint is stronger so it is useful to test the VHECR-induced cascade hypothesis including a scenario where VHECRs from extreme TeV blazars significantly contribute to the VHE DGB.  

\begin{figure*}[bt]
\begin{minipage}{0.49\linewidth}
\begin{center}
\includegraphics[width=\linewidth]{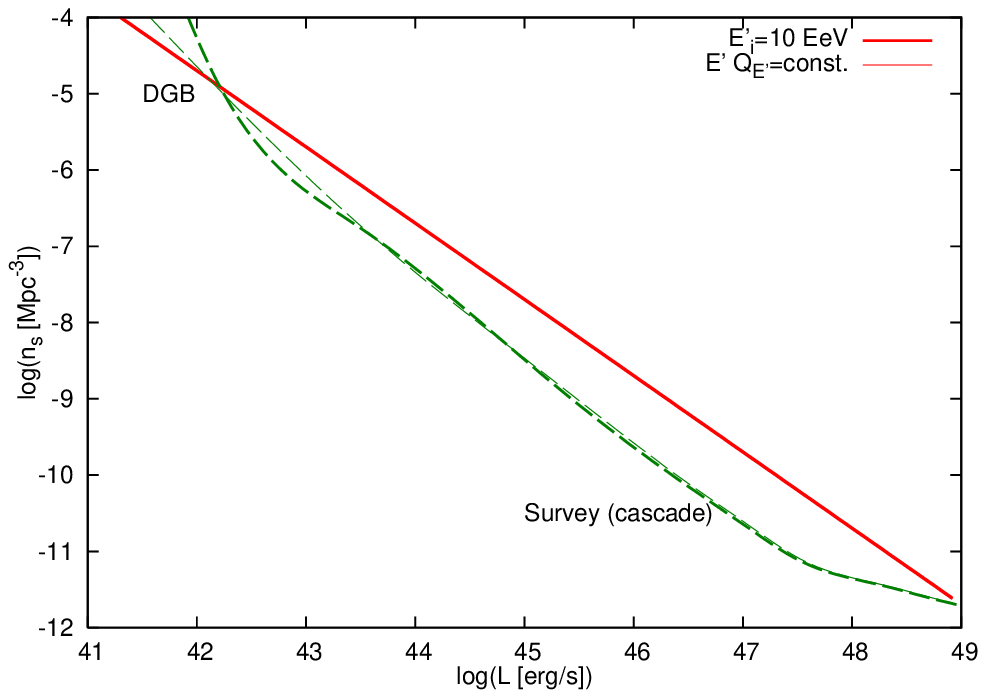}
\caption{Constraints from the DGB measured by \textit{Fermi} and from TeV surveys with future CTA-like IACTs.  For the thick curves, the near-mono-energetic gamma-ray injection over a half decade in logarithmic energy (with the central value of $E'_{\rm max}=10$~EeV) is assumed.  For the thin curves, $E' Q_{E'} =$~const. with ${E'}_{\rm min}={10}^{2.75}$~GeV and ${E'}_{\rm max}={10}^{11.25}$~GeV is assumed.}
\end{center}
\end{minipage}
\begin{minipage}{.05\linewidth}
\end{minipage}
\begin{minipage}{0.49\linewidth}
\begin{center}
\includegraphics[width=\linewidth]{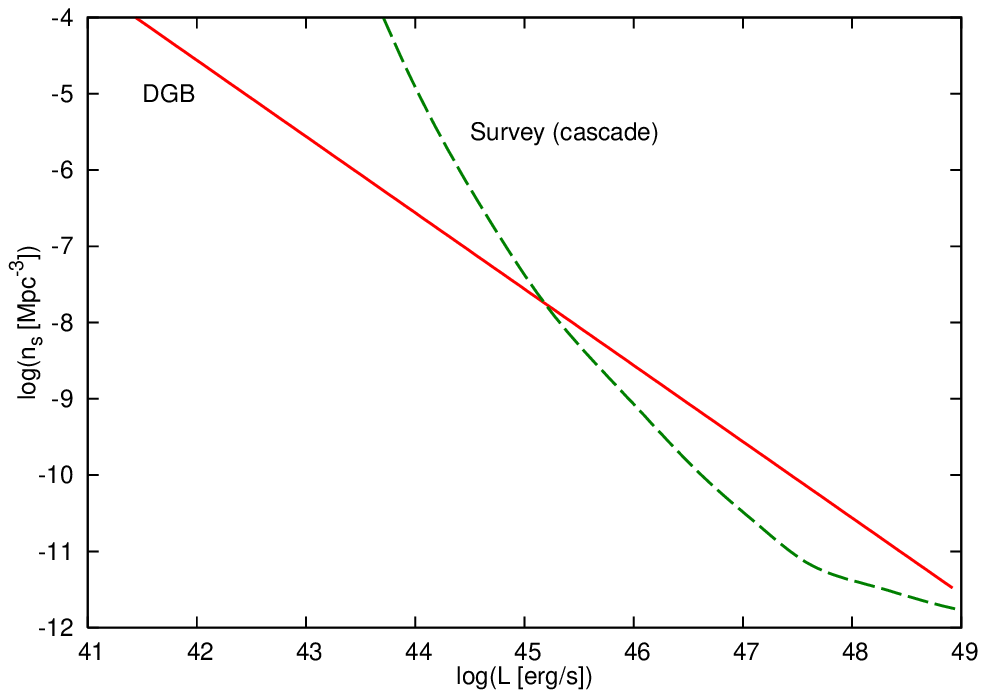}
\caption{Constraints from the DGB measured by \textit{Fermi} and TeV surveys with future CTA-like IACTs.  Here $L_{\rm vhecr}$ is used, and $q=2.6$ and ${E'}_{\rm max}={10}^{19}$~eV are assumed for the primary proton spectrum.  Survey (cascade) represents the case where the cascade emission contributes to the individual source flux, while DGB shows the constraint from the DGB fit obtained by \textit{Fermi}.}
\end{center}
\end{minipage}
\end{figure*}

As discussed in Section~3, UHE gamma-ray sources may also be high-energy neutrino sources, since UHE gamma rays should be produced by hadronic mechanisms.  Although the ratio of the neutrino flux to the gamma-ray flux strongly depends on source models, it is interesting to see the neutrino constraint for comparison.  In Figure~19, we show neutrino constraints obtained by IceCube-40~\cite{abb+11,spi11}.  For demonstrative purposes, we only consider a mono-energetic injection spectrum.  The point source limit is obtained for a mono-energetic spectrum with $E'_\nu = (1+z) E_\nu=1$~PeV.  On the other hand, the constraint from the background observation is always stronger, which is consistent with previous studies~\cite{lip06,sb10}. 
One sees that the neutrino constraints are more powerful than the gamma-ray constraint, if $E'_\nu Q_{E'_\nu} \sim E'_\gamma Q_{E'_\gamma}$ is satisfied.  
For blazars whose typical density is $n_s \sim {10}^{-7}~{\rm Mpc}^{-3}$, some hadronic blazar models need $L_\nu \sim L_\gamma \sim {10}^{45}-{10}^{46}~{\rm erg}~{\rm s}^{-1}$~\citep[e.g.,][]{ad03}.  Then, the neutrino limit of $E_\nu L_{E_\nu} \lesssim {10}^{43.5}~{\rm erg}~{\rm s}^{-1}$ suggests that optimistic models~\citep[e.g.,][]{man+01} have already been excluded~\cite{sb10}.  This demonstrates the importance of neutrino observations as a probe of UHECR accelerators.

\subsection{Detectability of individual sources with future TeV observations}
As seen before, the background observations by \textit{Fermi} and IceCube give us constraints on $n_s$ and $L$ (or $E L_E$).  And then one can estimate the expected number of VHE/UHE sources that are detectable by TeV surveys with future IACTs for given $n_s$ and $L$.  
Given the sensitivity with $F_E^{\rm lim}$, the number of detectable sources is estimated by
\begin{eqnarray}
N(L) = \frac{c}{H_0} \int^{z_{\rm lim}} dz \,\, \frac{{\Delta \Omega}_{\rm sur} d_L^2 n_s(z)}{{(1+z)}^2 \sqrt{\Omega_\Lambda+{(1+z)}^3 \Omega_m}} = \frac{n_s}{n_c},  
\end{eqnarray}
where $n_s$ is constrained from the DGB constraints discussed in Section~3. 
One should keep in mind that this estimate is valid only in the continuous limit.  We cannot exclude the existence of some outliers (accidentally nearby sources) located within $d_L^{\rm lim}$, and such sources might be seen even if $N(L) \lesssim 1$.   

Our results can be used for estimating the \textit{typical} maximum number of VHE gamma-ray sources with very hard gamma-ray spectra, assuming the next-generation IACTs such as CTA.  Here we mean that we do not consider the existence of outliers (accidentally nearby sources) located within $d_L^{\rm lim}$ by \textit{typical}.  The analytical estimate roughly gives 
\begin{eqnarray}
N(L)& \lesssim& 3~{\left( \frac{L}{{10}^{43}~{\rm erg}~{\rm s}^{-1}}  \right)}^{1/2} {\left( \frac{E {\mathcal G}_{E}}{0.2}  \right)}^{3/2}  {\left( \frac{t_{\rm sur}}{\rm 250~hr} \right)}^{3/4} \nonumber \\ 
&\times& {\left( \frac{E F_E^{50}}{{10}^{-13}~{\rm erg}~{\rm cm}^{-2}~{\rm s}^{-1}}  \right)}^{-3/2} {\left( \frac{{\Delta \Omega}_{\rm sur}}{4000~{\rm deg}^2}  \right)}^{1/4} 
{\left( \frac{{\Delta \Omega}_{\rm fov}}{{20}~{\rm deg}^2}  \right)}^{3/4}, 
\end{eqnarray}
though it is not accurate at large luminosities because the spectral suppression by the EBL is relevant for distant sources.  
The numerical results are shown in Figure~20.  In the cascade case, the typical number of detectable sources is expected to be at most $\sim 2-3$ with a 250~hr survey by CTA.  But this number can be increased to $\sim 7-8$ with $t_{\rm sur}=1200$~hr.  One obtains $\sim 3-8$, at $L_\gamma \sim {10}^{44}-{10}^{46}~{\rm erg}~{\rm s}^{-1}$ that is typical for high-peak BL Lac objects.  Note that the results are not sensitive to ${E'}_{\gamma}^{\rm max}$, once ${E'}_{\gamma}^{\rm max}$ is high enough for the cascaded gamma-ray spectrum to be similar. 
In the attenuation case, except for nearby outliers, it may be difficult to find such very hard gamma-ray emitters with CTA-like surveys, and the results highly depend on ${E'}_{\gamma}^{\rm max}$.    

\begin{figure*}[bt]
\begin{minipage}{0.49\linewidth}
\begin{center}
\includegraphics[width=\linewidth]{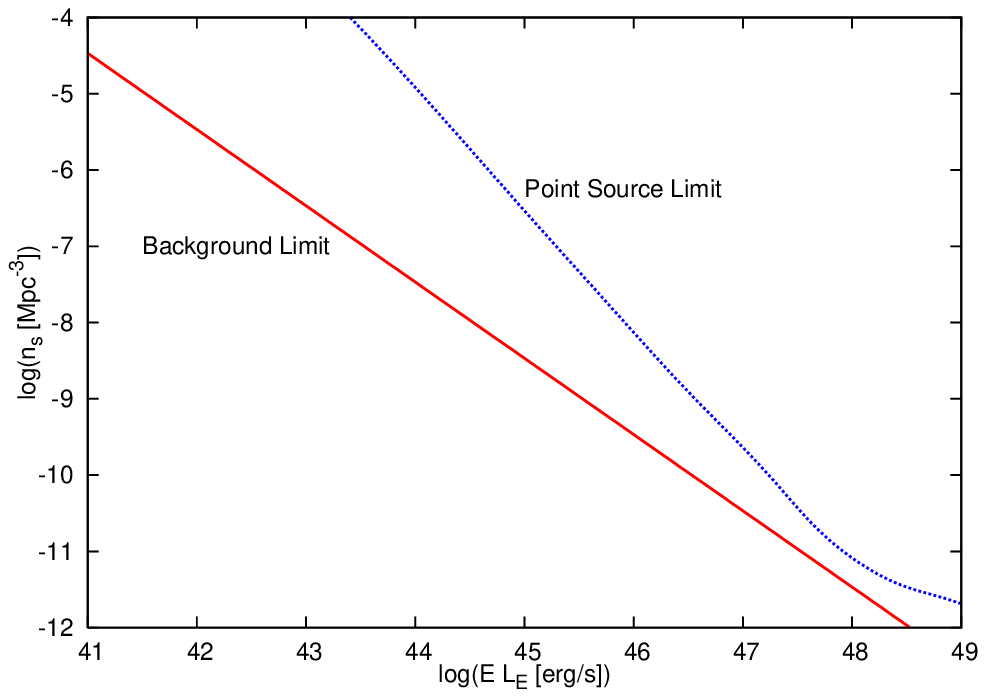}
\caption{Constraints from IceCube-40 observations of the cumulative neutrino background and point source observations.  The background limit is obtained from the quasi-differential sensitivity at $E=1$~PeV.   The point source limit is estimated from the quasi-differential sensitivity for the declination angle of $+30$~deg for the mono-energetic neutrino injection at ${E'}_i=1$~PeV.
\newline \, \newline}
\end{center}
\end{minipage}
\begin{minipage}{.05\linewidth}
\end{minipage}
\begin{minipage}{0.49\linewidth}
\begin{center}
\includegraphics[width=\linewidth]{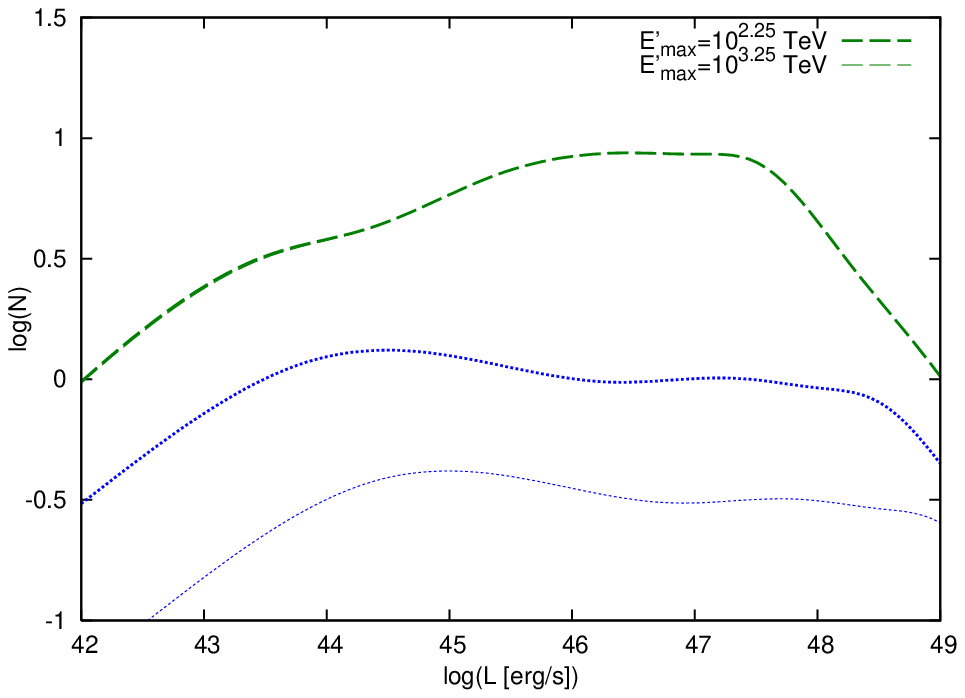}
\caption{Estimates on the typical maximum number of VHE gamma-ray sources that are allowed by the DGB measurement and are detectable by TeV surveys with future CTA-like IACTs.  The photon index is set to $s=1.5$ with ${E'}_{\rm max}={10}^{2.25}$~TeV or ${E'}_{\rm max}={10}^{3.25}$~TeV.   The dashed curves represent the case where the cascade emission contributes to the individual source flux, whereas the dotted curves are for the case where only the attenuated emission is relevant.}
\end{center}
\end{minipage}
\end{figure*}

Next, we consider UHE gamma-ray emitters, motivated by searches for the origin of UHECRs.  
For demonstration, we consider a near-mono-energetic neutrino/gamma-ray emitter, where UHE photon and neutrinos are comparably produced with $E'_{\nu} Q_{E'_{\nu}} = E'_{\gamma} Q_{E'_\gamma}$ at $10$~EeV (cf. Figure~17).  Then, the IceCube sensitivity on the neutrino background at $10$~EeV gives an estimate on the typical maximum number of UHE gamma-ray emitters, as shown in Figure~21.  Based on IceCube-40 limits on the neutrino background, one sees that $\gtrsim 10$ distant and bright UHE gamma-ray emitters may be detected for $t_{\rm sur}=1200$~hr.  However, by the time when CTA starts gamma-ray observations, IceCube will have improved sensitivities to the cumulative neutrino background.  Therefore, though IceCube may eventually detect extragalactic neutrino sources, we also use the full IceCube background limit to estimate the typical number of detectable UHE gamma-ray emitters.  In this case, one sees that $\sim 2-3$ distant and luminous UHE gamma-ray emitters may be detected.  Note that the IceCube sensitivity is better at $\sim$~PeV than at $\sim$~EeV.  

Just for demonstrative purposes, we also show the case for $E'_{\nu} Q_{E'_{\nu}} = E'_{\gamma} Q_{E'_\gamma}=$const., where the expected number of detectable VHE/UHE gamma-ray emitters is smaller than unity because of the stricter neutrino limit.  In this case, we have
\begin{eqnarray}
N(L) &\lesssim& 0.3~{\left( \frac{L}{{10}^{43}~{\rm erg}~{\rm s}^{-1}}  \right)}^{1/2} {\left( \frac{E {\mathcal G}_{E}}{0.2}  \right)}^{1/2} {\left( \frac{\mathcal R}{20}  \right)}^{-1}  {\left( \frac{t_{\rm sur}}{\rm 250~hr} \right)}^{3/4} \nonumber \\ 
&\times& {\left( \frac{E F_E^{50}}{{10}^{-13}~{\rm erg}~{\rm cm}^{-2}~{\rm s}^{-1}}  \right)}^{-3/2} {\left( \frac{{\Delta \Omega}_{\rm sur}}{4000~{\rm deg}^2}  \right)}^{1/4} 
{\left( \frac{{\Delta \Omega}_{\rm fov}}{{20}~{\rm deg}^2}  \right)}^{3/4}, 
\end{eqnarray}
where ${\mathcal R} = {\rm ln} ({E'}_{\gamma}^{\rm max}/{E'}_{\gamma}^{\rm min})$ and Eq.~(4.16) also agrees with the numerical result at low luminosities, $L_{\gamma} \lesssim {10}^{43.5}~{\rm erg}~{\rm s}^{-1}$.  
This demonstrates that neutrino constraints are more important than gamma-ray constraints for gamma-ray emitters with $E'_{\nu} Q_{E'_{\nu}} \gtrsim E'_{\gamma} Q_{E'_\gamma}$.  Such cases are realized, e.g., when gamma rays are dominantly produced by photomeson production and/or the $pp$ reaction.  

Finally, we consider the case of the VHECR-induced cascade. The result is shown in Figure~22. 
For VHECR luminosities of $L_{\rm vhecr} \sim {10}^{45}-{10}^{46}~{\rm erg}~{\rm s}^{-1}$ that are typically required in the VHECR-induced cascade scenario for extreme TeV blazars~\cite{ess+11,mur+12,raz+12}, it is possible to find a couple of such VHECR sources with future TeV surveys.   If more dedicated surveys are possible or the integrated sensitivity is used, we expect that more VHECR-induced cascade sources can be seen if they significantly contribute to the VHE DGB.  Also, there may accidentally be nearby sources. 
We emphasize that, in any case, it is important for future IACT surveys to search for such very hard gamma-ray sources.  Those sources do not have to be detected by \textit{Fermi} because the GeV flux may be much lower than the TeV flux due to their very hard spectra, because the IACTs have better sensitivities than \textit{Fermi}.     
Of course, multi-wavelength observations are definitely relevant, since sources that are discovered at other wavelengths may eventually be confirmed as gamma-ray sources.  However, one should also keep in mind that the connection between low energies and VHE/UHE is not clear in the presence of contributions produced by CRs.  

\begin{figure*}[bt]
\begin{minipage}{0.49\linewidth}
\begin{center}
\includegraphics[width=\linewidth]{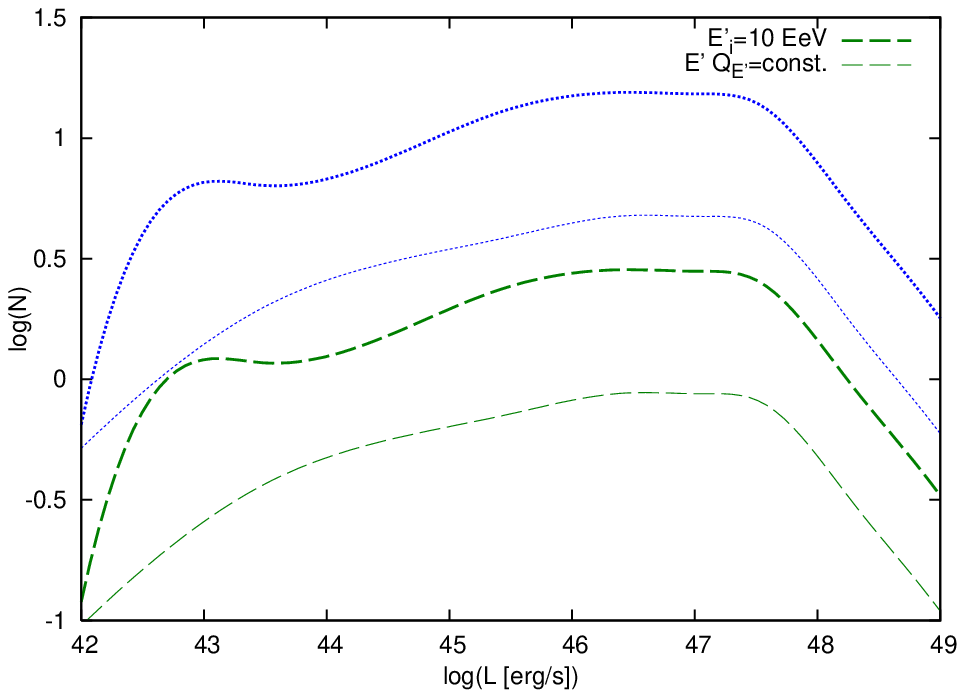}
\caption{Estimates on the typical maximum number of steady UHE gamma-ray sources that are allowed by the neutrino background observation and are detectable by TeV surveys with future CTA-like IACTs.  The dashed curves are for the future full IceCube limits, while the dotted curves are for the present IceCube-40 limits.  For the thick curves, the near-mono-energetic UHE injection over a half decade in logarithmic energy (with the central value of $10$~EeV) is assumed (for both gamma rays and neutrinos).  For the thin curves, $E' Q_{E'} =$~const. with ${E'}_{\rm min}={10}^{2.75}$~GeV and ${E'}_{\rm max}={10}^{11.25}$~GeV is assumed (for both gamma rays and neutrinos).  The limits are obtained assuming $E'_{\nu} Q_{E'_{\nu}} = E'_{\gamma} Q_{E'_\gamma}$.}
\end{center}
\end{minipage}
\begin{minipage}{.05\linewidth}
\end{minipage}
\begin{minipage}{0.49\linewidth}
\begin{center}
\includegraphics[width=\linewidth]{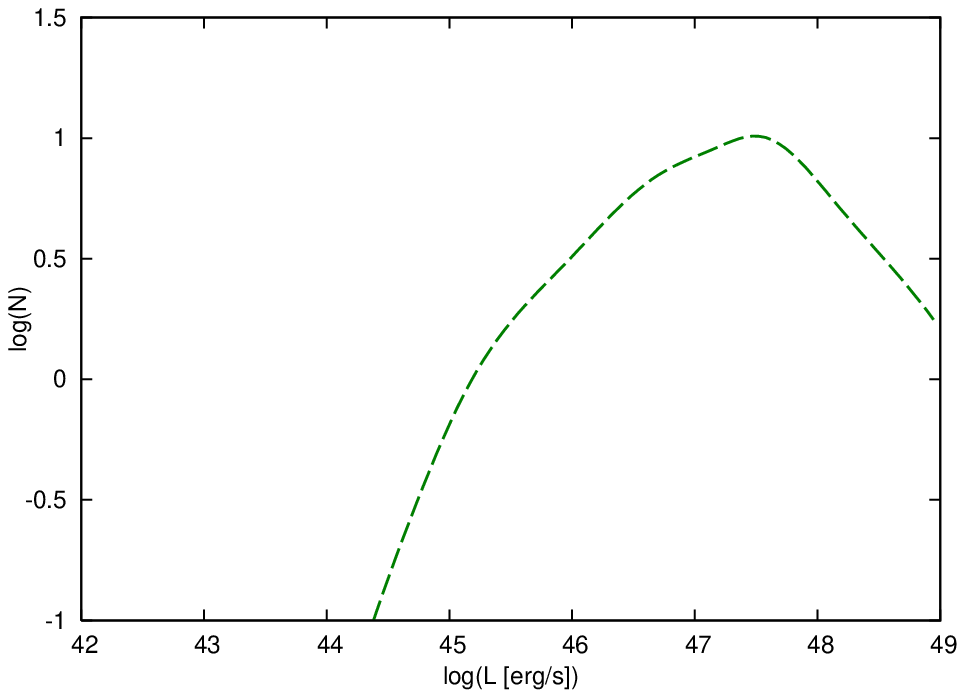}
\caption{The estimate on the typical maximum number of VHECR-induced cascade sources that are allowed by the DGB measurement and are detectable by TeV surveys with future CTA-like IACTs.  The CR index is set to $q=2.6$ with ${E'}_{\rm max}={10}^{19}$~eV.  The dashed curve represents the case where the cascade emission contributes to the individual source flux.
\newline \, \newline \, \newline \, \newline \, \newline \, \newline \, \newline}
\end{center}
\end{minipage}
\end{figure*}

The results presented in this section are general since we have not specified VHE/UHE sources. 
But, in the case of steady extragalactic sources, as far as we know, the main applicable VHE/UHE sources will be blazars.  Interestingly, we may already have seen such intergalactic cascade sources.  As explicitly shown in Murase et al.~\cite{mur+12} and discussed in Section~2, both the gamma-ray-induced cascade and the CR-induced cascade can explain VHE spectra observed in extreme TeV blazars.  So far, the number of such extreme sources is only a few, but we may find more such sources in the near future.  The cascade hypothesis for the VHE DGB, suggested in Section~2, could be consistent with the existence of those extreme TeV blazars showing very hard spectra.   

Note that we have considered steady sources.  For transient sources, the strategy is different and all-sky monitors such as HAWC start to play an important role. 
In the case of AGN, for example, they often have flaring activities, and they show the high state in some episodes.  Then, follow-up observations IACTs such as CTA should be useful to reveal their high-energy behaviors.  In particular, even cascade signals may lead to long-term transients called pair echoes~\cite{mur+08b,mur12}, which can be very important for follow-up observations.

\section{Summary and discussion}
In this paper, we focus on the new information from two types of searches that seem to be close to theoretically-predicted fluxes: cascaded gamma rays and neutrinos.  When VHE/UHE gamma rays undergo pair production, the original gamma rays are lost.  However, the electrons and positrons lose energy in part by IC scattering, producing regenerated, lower-energy gamma rays.  The repetition of these processes produces electromagnetic cascades.  If sufficiently high-energy gamma rays are emitted, through energy conservation, they are cascaded down in the nominal detection range of \textit{Fermi}, thus revealing higher-energy sources indirectly. 
With detailed numerical calculations, we demonstrated the potential importance of contributions from the cascades induced by high-energy gamma rays and VHECRs/UHECRs. 
 
The origin of the DGB is largely unknown, so that new approaches are needed.  
Gamma-ray cascades provide a test for high-energy injections and may give a maximum extragalactic contribution to the VHE DGB, for which there are hints in the VHE DGB.  Neutrinos are also useful as another test if the emission is hadronic, and those can be detected directly, with their emitted energies affected only modestly by the redshift. 
Together, these gamma-ray and neutrino probes can reveal the existence and nature of high-energy sources, whether astrophysical (e.g., CR accelerators) or exotic (e.g., dark matter annihilation or decay).  They are complementary to studies on the population of sources and the anisotropy in the DGB.  Our results can be summarized as follows.

(1) The EBL attenuation is typically relevant at $\gtrsim 100$~GeV energies, though quantitative details depend on redshift evolution models and EBL models.  
If the VHE DGB is the case, we would have three possibilities: (a) extragalactic origins with a ``VHE Excess" in the $\sim 0.1-10$~TeV range of the effective primary gamma-ray spectrum, (b) extragalactic origins with significant cascade contributions, (c) Galactic origins or other possibilities involving exotic physics such as Lorentz-invariance violation.  It is natural to expect gamma-ray injections at sufficiently high energies, where a cascade component should start to dominate over the EBL-attenuated gamma-ray component attributed to the lower-energy DGB if sources with hard spectra have enough energy budgets.  In such a cascade scenario (b), slower redshift evolution models are better to obtain harder DGB spectra, though fast redshift evolution models are required for the VHECR-indued cascade to avoid the overshooting problem that the required CR energy budget violates the observed CR flux.  
The no redshift evolution case can give a reasonable cascade upper limit on the VHE DGB although the argument depends on details of the injection.  
The precise determination of the VHE DGB will enable us to test the cascade hypothesis for the VHE DGB and fast evolution models may be excluded.  
Compared to VHECR-induced cascade sources, gamma-ray-induced cascade sources may help get harder DGB spectra just below TeV energies, and the discrimination between the two is in principle possible by upcoming precise measurements of the VHE DGB if the EBL is well-determined.  But the prediction of the VHE DGB is affected by the EBL.  Though the VHE DGB at $\gtrsim 1$~TeV would become difficult to measure with existing telescopes, it is useful to test extragalactic scenarios for the VHE DGB and constraining the EBL.  
Anisotropy signals are also expected to change around the transition energy, where cascade components become dominant over other attenuation components from unresolved point sources, though details depend on the void IGMF.   If the IGMF is not too small, resolving more point sources in the near future would be relevant since the cascade component would remain unresolved, making the DGB relatively harder in the VHE range.     

(2) Another test of the cascade hypothesis for the VHE DGB can be achieved by future TeV surveys of individual distant sources with CTA.  Weak void IGMFs with $\lesssim {10}^{-15}-{10}^{-14}$~G are motivated by intergalactic cascade scenarios for extreme TeV blazars such as 1ES 0229+200, although predictions depend on IGMFs.  Hence, the intergalactic cascade hypothesis for the VHE DGB may be consistent with these cascade scenarios for individual TeV sources. But the former can be viable even if the latter is excluded.  
Even if the maximum typical number of detectable sources of cascaded gamma rays is at most $\sim 10$, confirming or constraining these cascade signatures is crucial in order to understand the origin of the DGB and physics of VHE/UHE sources. Possibly, we may obtain important clues to extragalactic VHECR/UHECR accelerators or sources, which can also help know the origin of CRs.    

More conservatively, we utilized the DGB just in order to place constraints on the energy budget of gamma rays, and obtained general cascade constraints on the gamma-ray energy budget.  
They can be compared to IceCube constraints on the neutrino energy budget, and we showed that  
neutrino observations are indeed powerful as a probe of VHE/UHE hadronic sources with $E_\nu Q_{E_\nu} \sim E_\gamma Q_{E_\gamma}$.  We also demonstrated that future TeV surveys by next IACTs such as CTA are useful for identifying CR accelerators, and UHE gamma-ray and/or VHECR/UHECR emitters might be seen as cascaded TeV gamma-ray sources.  

Our results strengthen the importance of observing the VHE DGB and the neutrino background, and support the case where current and future multi-messenger searches can play important roles in revealing extragalactic VHE/UHE sources, including CR accelerators.   

After our paper was submitted and posted on arXiv (arXiv:1205.5755), a related paper~\cite{ii12} came out.  Our work is general and comprehensive in the sense that we derived multi-messenger constraints on the energy budget and addressed future prospects of detecting individual sources as well as investigated general cascade contributions to the VHE DGB.  Our numerical method is even applicable to the Klein-Nishina cascade induced by UHE gamma rays and the VHECR-induced cascade as well as the Thomson cascade induced by VHE photons. 

\section*{Acknowledgments} 
K. M. is supported by JSPS and CCAPP.  The research of J. F. B. is supported by NSF Grant PHY-1101216. H. T. is supported by JSPS. 
We thank an anonymous referee for many valuable comments.  We also thank Markus Ackermann, Markus Ahlers, John Cairns, Paolo Coppi, Charles Dermer, Shunsaku Horiuchi, Kunihito Ioka, and Alexander Kusenko for helpful discussions. 

\newpage

\bibliographystyle{JHEP}
\bibliography{ms}


\end{document}